\documentclass[
  twocolumn,prl,aps,longbibliography,nofootinbib
]{revtex4-2}

\usepackage[utf8]{inputenc}
\usepackage[T1]{fontenc}
\usepackage{amsmath,amssymb,amsfonts,mathtools,bm}
\usepackage{graphicx}
\usepackage{booktabs}
\usepackage{enumitem}
\usepackage[hidelinks]{hyperref}
\usepackage{cleveref}
\usepackage{soul}
\usepackage{xcolor}
\usepackage[most]{tcolorbox}

\newcommand{\E}{\mathbb{E}}

\newcommand{\id}{\mathbb{I}}
\newcommand{\Tr}{\operatorname{Tr}}
\newcommand{\Var}{\operatorname{Var}}
\newcommand{\Cov}{\operatorname{Cov}}
\newcommand{\supp}{\operatorname{supp}}
\newcommand{\wt}{\operatorname{wt}}
\newcommand{\Pset}{\mathcal{P}_{\le K}}
\newcommand{\abs}[1]{\left\lvert #1\right\rvert}
\newcommand{\norm}[1]{\left\lVert #1\right\rVert}
\newcommand{\muhat}{\widehat{\mu}}
\newcommand{\thetahat}{\widehat{\theta}}
\newcommand{\ket}[1]{\left\lvert #1\right\rangle}

\newtcolorbox{changed}{breakable,enhanced,colback=yellow!45,colframe=yellow!45,boxrule=0pt,arc=0pt,boxsep=1pt,left=1pt,right=1pt,top=1pt,bottom=1pt,before skip=2pt,after skip=2pt}

\begin{document}

\title{Quantum-enhanced metrology with classical shadows}
\author{Luca Ion}\email{luca.ion@uv.es}
\author{Manuel Gessner}\email{manuel.gessner@uv.es}
\affiliation{Instituto de Física Corpuscular (IFIC), CSIC-Universitat de València and Departament de Física Teòrica, UV, Av. Vicent Andrés Estellés, 19, E-46100 Burjassot (València), Spain}

% \begin{abstract}
% Quantum metrology often relies on measuring carefully designed observables that provide maximal information about the parameter of interest, but these observables may be hard to access or even theoretically unknown. We introduce a shadow-native protocol that uses randomized Pauli measurements to access many observables simultaneously and combines their calibrated local response into an optimized method-of-moments estimator. For local Pauli classical shadows, we derive the exact estimator variance, conservative high-probability sample-complexity guarantees, and shadow information bounds. The protocol is especially suited to parameters encoded across many low-weight Pauli observables, as illustrated in several examples. Our work establishes randomized Pauli measurements as a setting-agnostic and broadly applicable strategy for quantum parameter estimation.\mg{reframe: add more results, e.g. entanglement enhanced performance}
% \end{abstract}

\begin{abstract}
Quantum metrology relies on observables that maximize information about the parameter of interest, but are often hard to access. We develop a protocol that combines Pauli
moments estimated from a fixed ensemble of randomized local
measurements into an optimized method-of-moments estimator. For local Pauli classical shadows, we derive the shadow information which determines estimator variance and sample-complexity guarantees. Identifying the separable-state sensitivity limit, we show that entanglement can lead to a constant-factor sensitivity enhancement, as illustrated with one-axis-twisted probes. This protocol allows for efficient circuit-based metrology without choosing an observable.
\end{abstract}

\maketitle

\textit{Introduction.---}Quantum metrology provides a framework for estimating unknown parameters encoded in quantum states with the highest precision allowed by quantum mechanics~\cite{GiovannettiSCIENCE2004,GiovannettiPRL2006,GiovannettiNATUREPHOTONICS2011,PezzeRMP2018}. In its most powerful formulation, it identifies the ultimate asymptotic precision limit for any locally unbiased estimator and determines measurements that attain this limit~\cite{HelstromBOOK1976,HolevoBOOK1982,BraunsteinPRL1994}. Determining these measurements generally requires detailed
knowledge of the parameter-dependent state. In high-dimensional settings this is often challenging or even unfeasible, and the resulting observable may have a complex structure that is inaccessible in practice.

An alternative route is provided by method-of-moments estimation. Instead of optimizing over all quantum measurements, one starts from an accessible observable and infers the parameter from its calibrated response near a working point. This idea is at the heart of error-propagation formulae and spin-squeezing parameters, which are widely used in experiments because of their practicality when employed with linear observables, even when they are not globally optimal~\cite{WinelandPRA1992,KitagawaPRA1993,MaPHYSREP2011,PezzeRMP2018}. More general moment-based criteria based on nonlinear observables can reveal better parameter sensitivity beyond standard linear squeezing parameters, taking advantage of the high sensitivity of non-Gaussian quantum states~\cite{GessnerPRL2019}. Nevertheless, optimal nonlinear observables are also hard to access and are typically reconstructed from a large number of independently measured observables~\cite{XuPRL2022}.

Classical shadows provide a way to reuse randomized measurements for estimating many properties of a quantum state without performing full state tomography~\cite{HuangNATUREPHYSICS2020}. A single randomized data set can be post-processed to predict many target expectation values, and for suitable measurement ensembles the required number of shots scales only logarithmically with the number of targets and polynomially with an observable-dependent shadow norm~\cite{HuangNATUREPHYSICS2020}. More broadly, randomized measurements form a versatile toolbox in quantum information processing, with applications to state characterization~\cite{StruchalinPRXQUANTUM2021}, fidelity estimation and nonlinear functions~\cite{HuangNATUREPHYSICS2020,AcharyaPRA2021}, entanglement and correlation diagnostics~\cite{ElbenPRL2020,ElbenNATUREREVIEWS2023}, and metrological sensitivity~\cite{RathPRL2021,YuPRRESEARCH2021,ZhangPRL2025}.  Their ability to reuse measurement data multiple times makes shadows particularly powerful when the relevant information is distributed over many accessible observables. More recent work has started to explore shadows in multiparameter estimation using global randomized
measurements~\cite{ZhouPRXQUANTUM2026} and few-qubit
readout~\cite{MaoARXIV2026}.

Here we show how local Pauli classical shadows enable optimized
method-of-moments quantum parameter estimation without tailoring
the measurement to the probe or encoding. A single randomized
dataset provides estimates of all Pauli strings up to a chosen
maximum weight, each contributing calibrated local information
about the unknown parameter. Generalized least squares combines
these correlated moment estimates into a locally unbiased
estimator with optimal linear weighting, extracting parameter
sensitivity without identifying or directly measuring the optimal
metrological observable. 

For local Pauli shadows, we derive the exact covariance
matrix of the estimated Pauli moments and introduce the
shadow information, whose inverse determines the variance
of the optimized estimator per measurement shot as well as Bernstein
concentration bounds and high-probability sample-complexity
guarantees. We establish relations between the shadow information and established metrological sensitivity measures such as generalized squeezing
parameters~\cite{GessnerPRL2019} and the quantum Fisher information~\cite{BraunsteinPRL1994}. Upper bounds on the shadow
information for separable and arbitrary probe states identify the classical and quantum precision limits for this protocol and provide a sensitivity 
threshold for witnessing metrologically useful probe entanglement. Our results show that randomized Pauli measurements can serve as a setting-agnostic strategy for quantum parameter estimation, replacing tailored measurement design by universal data acquisition and optimized classical post-processing.
% Resolved: \mg{don't or only as a side remark discuss direct strategy here}

\textit{Quantum parameter estimation from low-weight Pauli strings.---}The goal of quantum metrology is to extract an unknown parameter encoded in a quantum state as efficiently as possible, typically by performing a carefully chosen quantum measurement whose outcomes vary sensitively with the parameter. Here we adapt this principle to a setting in which many simple, generic observables are measured simultaneously and their individual parameter responses are combined through optimized classical post-processing. We focus on qubit systems, as naturally realized in quantum-circuit models, where Pauli strings form a natural family of accessible observables. A Pauli string is a tensor product of local Pauli operators, $P=P_1\otimes\cdots\otimes P_N$, where $P_j\in\{\mathbb I,X,Y,Z\}$. Its support $\supp(P)=\{j:P_j\neq\mathbb I\}$ is the set of sites on which it acts nontrivially, and its weight is $\wt(P)=|\supp(P)|$.

Specifically, we consider an unknown parameter $\theta$ encoded in an $N$-qubit state $\rho_\theta$ and use the expectation values of all nonidentity Pauli strings up to weight $K$, collected in the set $\Pset=\{P:1\le \wt(P)\le K\}$. The corresponding calibration data are summarized by the moment vector $\boldsymbol{\mu}(\theta)=(\mu_P(\theta))_{P\in\Pset}$, with $\mu_P(\theta)=\Tr(P\rho_\theta)$ and a total number of $M=|\Pset|=\sum_{k=1}^{K}{N\choose k}3^k$ observables.

In the local-estimation paradigm, the unknown value of $\theta$ is assumed to lie in a small neighbourhood of a known working point $\theta_0$. In the inference stage, $n$ independent measurement shots provide an empirical estimator $\boldsymbol{\muhat}$ of $\boldsymbol{\mu}(\theta)$. Comparing this estimate with the calibrated response curve, expanded locally as $\boldsymbol{\mu}(\theta)=\boldsymbol{\mu}(\theta_0)+\boldsymbol{s}(\theta-\theta_0)+O((\theta-\theta_0)^2)$, gives an estimate of the deviation of $\theta$ from $\theta_0$, where $\boldsymbol{s}=\partial_\theta\boldsymbol{\mu}(\theta)|_{\theta_0}$ is the local response vector.

To infer an unknown value of $\theta$ from these moments, we assume that the calibration curve $\boldsymbol{\mu}(\theta)$ and response vector $\boldsymbol{s}$ are known in a neighbourhood of the working point, either through prior experiments or from a trusted theoretical model. Following the standard treatment of local quantum metrology, we treat this calibration as a prerequisite and exclude its cost from the measurement budget for the final estimation. Experimentally, calibration through the linearized expansions of $\boldsymbol{\mu}(\theta)$ requires estimating Pauli expectation values $\boldsymbol{\mu}(\theta_0)$ and their local responses $\boldsymbol{s}$ at the working point---tasks routinely encountered in variational quantum algorithms, for which established measurement-grouping \cite{Verteletskyi_2020} and gradient-estimation techniques \cite{PhysRevA.99.032331} provide practical approaches.

For known calibration and covariance, generalized least squares
(GLS) minimizes the variance among locally unbiased 
estimators constructed from the selected moment estimates.
At the working point, the minimum variance is
\begin{align}
    \Var_{\theta_0}(\thetahat_{\rm GLS})
    =
    \frac{1}{n}
    \frac{1}{\boldsymbol{s}^T\Sigma^{-1}\boldsymbol{s}} .
    \label{eq:var_gls}
\end{align}
Here $\Sigma$ denotes the single-shot covariance matrix of the moment estimators at $\theta_0$ and $n$ is the number of shots. The derivation is given in the Supplemental Material~\cite{SM}. The calibrated response vector $\boldsymbol{s}$ describes how
the moments change with the parameter, while $\Sigma$ captures
their fluctuations and correlations for the chosen state and
measurement protocol. We now derive $\Sigma$ for shadow-based estimators.

\textit{Classical-shadow estimator.---}We construct a locally unbiased moment estimator from classical shadows, using the local Pauli-shadow protocol, in which each qubit is measured in an independently and uniformly chosen Pauli basis. For each shot $\ell=1,\ldots,n$, the state $\rho_\theta$ is prepared, a local Pauli basis $B^{(\ell)}=B_1^{(\ell)}\otimes\cdots\otimes B_N^{(\ell)}$ is sampled uniformly with $B_j^{(\ell)}\in\{X,Y,Z\}$, and the corresponding outcome $\boldsymbol{b}^{(\ell)}=(b_1^{(\ell)},\ldots,b_N^{(\ell)})$, with $b_j^{(\ell)}\in\{\pm1\}$, is recorded.

For a target Pauli string $P$, the single-shot shadow estimator of $\mu_P(\theta)$ is~\cite{HuangNATUREPHYSICS2020,HadfieldCMP2022}
\begin{align}
  X_P^{(\ell)}
  =
  \begin{cases}
  3^{\wt(P)}
  \displaystyle\prod_{j\in\supp(P)} b_j^{(\ell)}, & P\sim B^{(\ell)},\\
  0, & P\not\sim B^{(\ell)},
  \end{cases}
  \label{eq:shadow-single}
\end{align}
Here $P\sim Q$ means that $P_i=Q_i$ on every site where both strings are nonidentity; since $B^{(\ell)}$ has no identity factors, $P\sim B^{(\ell)}$ means that the sampled basis matches $P$ on $\supp(P)$.
%The factor $3^{\wt(P)}$ compensates for the probability $3^{-\wt(P)}$ of sampling the required local bases, so that $\mathbb E[X_P^{(\ell)}]=\mu_P(\theta)$. 

Stacking the empirical shadow estimates $\widehat{\mu}^{\rm cs}_P=(\sum_{\ell=1}^{n}X_P^{(\ell)})/n$ for all $P\in\Pset$ gives the vector $\boldsymbol{\muhat}^{\rm cs}$ with covariance $\operatorname{Cov}(\boldsymbol{\muhat}^{\rm cs})=\Gamma/n$. For local Pauli shadows, the single-shot covariance matrix admits the closed form (see the Supplemental Material~\cite{SM})~\cite{HadfieldCMP2022,hayashi2026finitesampleselectedcovariancespectra}
\begin{align}
\Gamma_{PQ}
=
\begin{cases}
3^{|\supp(P)\cap\supp(Q)|}\langle PQ\rangle-\mu_P\mu_Q, & P\sim Q,\\
-\mu_P\mu_Q, & P\not\sim Q,
\end{cases}
\label{eq:shadow-cov-main}
\end{align}
where all expectation values are evaluated at the working point. The diagonal entries contain the Pauli-weight overhead, while the off-diagonal entries describe correlations generated by estimating many Pauli moments from the same randomized data.

Optimally combining the shadow moments with GLS yields the estimator variance of $\theta$ given in Eq.~(\ref{eq:var_gls}) with $\Sigma=\Gamma$:
\begin{align}
  \Var_{\theta_0}(\thetahat_{\rm cs})
  =
  \frac{1}{n\mathcal{I}[\rho]},
  \label{eq:shadow-var}
\end{align}
where we introduced the shadow information
\begin{align}
    \mathcal{I}[\rho]=\boldsymbol{s}^T\Gamma^{-1}\boldsymbol{s}.
\end{align}
This information function determines the precision of the estimator $\thetahat_{\rm cs}$ and depends on the fluctuations contained in the shadow covariance matrix, in direct analogy to generalized squeezing parameters in conventional direct-measurement metrology protocols~\cite{GessnerPRL2019}. Its inverse determines the asymptotic per-shot estimator variance.

The shadow information also enters the finite-sample guarantees. The sample complexity tells us how many measurements are needed for the estimator to achieve a certain accuracy $\eta$ with a certain success probability $1-\delta$. Bernstein concentration~\cite{BoucheronBOOK2013} gives the
finite-sample bound derived in the
Supplemental Material~\cite{SM}.
In the small-error regime, an estimation error
\begin{align}
    |\thetahat_{\rm cs}-\theta_0|\le\eta,
\end{align}
with failure probability at most $\delta$, is reached with a sample size of the order of
\begin{align}
  n
  \sim
  \frac{1}{\eta^2}
  \frac{1}{\mathcal{I}[\rho]}
  \log\frac{1}{\delta}.
  \label{eq:shadow-sample}
\end{align}
A simple worst-case sufficient bound for Pauli strings of weight at most $K$ is obtained by bounding the shadow information $\mathcal{I}[\rho]$ from below, and reads
\begin{align}
  n
  \gtrsim
  \frac{3^{2K}}{\eta^2}
  \frac{\|\boldsymbol{s}\|_1^2}{\|\boldsymbol{s}\|_2^4}
  \log\frac{1}{\delta}.
  \label{eq:suff-cs}
\end{align}
This bound is conservative, but it makes explicit that the random-basis overhead is controlled by the maximum Pauli weight.

The key advantage of shadow measurements is that all moments
entering $\mathcal I[\rho]$ are estimated from the same $n$ shots.
For comparison, a naive direct-measurement strategy consists in splitting these
shots equally among $M$ independently measured observables.
This yields a diagonal covariance matrix with moment variances
increased by a factor $M$, leading to substantially
larger estimation errors when $M$ is large (see the
Supplemental Material~\cite{SM} for this benchmark and grouped commuting-measurement protocols).

\textit{Upper bounds on the shadow information.---}We now identify limits on the information that is accessible through the fixed shadow readout.
Consider collective phase encoding
$\rho_\theta=e^{-i\theta H}\rho_0 e^{i\theta H}$ with
$H=J_{\boldsymbol n}=\frac12\sum_i\boldsymbol n\cdot\boldsymbol\sigma_i$
and $\norm{\boldsymbol n}=1$. In the following, all quantities are evaluated at the working point, and we
omit the $\theta$ subscript in $\rho_\theta$. For this encoding, the shadow information obeys the bound
derived in the Supplemental Material~\cite{SM},
\begin{align}
    \mathcal{I}[\rho]\leq\frac{2N\chi^{-2}[\rho]}{2N+\chi^{-2}[\rho]}\leq \chi^{-2}[\rho],
\end{align}
where 
\begin{align}
\chi^{-2}[\rho]=\frac{\left|\frac{\partial}{\partial
\theta}\langle A\rangle\right|^2}{\Var_{\rho}(A)},
\end{align}
with $A=\sum_{P\in\Pset}a_PP$ and $a_P$ form the vector $\bm{a}$ that identifies the coefficients of the GLS estimator, that is, it minimizes $(\bm{a}^T\Gamma\bm{a})/(\bm{a}^T\bm{s})^2$. 
% A well-known hierarchy in metrology~\cite{} further establishes that
% \begin{align}\label{eq:chi2F}
%     \chi^{-2}[\rho]\le F[\rho],
% \end{align}
% where $F[\rho]=\sum_{\lambda:p_{\lambda}>0}\frac{(\partial p_{\lambda})^2}{p_{\lambda}}$ with $p_{\lambda}=\Tr\{\rho \Pi_{\lambda}\}$ is the classical Fisher information obtained from measurements of the observable $A$ with spectral decomposition $A=\sum_{\lambda}\lambda \Pi_{\lambda}$.

At a fixed working point, the shadow information obtained from Pauli strings up to weight $K$ is given by $\mathcal I_K[\rho]=\boldsymbol s_K^T\Gamma_K^{-1}\boldsymbol s_K$. At the lowest order, $K=1$, we restrict to linear observables and the shadow information is limited by
\begin{equation}\label{eq:I1_bound}
    \mathcal{I}_1[\rho]\le N/2.
\end{equation}
Increasing $K$ expands the set of moments available to the
estimator. Since GLS can assign zero weight to any additional
moment, the optimized information is nondecreasing:
\begin{equation}\label{eq:hierarchyIK}
    \mathcal I_1[\rho]\le\cdots\le \mathcal I_N[\rho].
\end{equation}

The full sensitivity available in the complete record of measurement bases
and outcomes is characterized by its classical Fisher information. It bounds the
shadow information according to
\begin{equation}
 \mathcal{I}_N[\rho]\le F_{\mathrm{cs}}[\rho],
 \label{eq:shadow-information-hierarchy1}
\end{equation}
where
\begin{equation}
 F_{\mathrm{cs}}[\rho]
 =\sum_{\omega:p_\omega>0}
 \frac{(\partial_\theta p_\omega)^2}{p_\omega}.
\end{equation}
Here $\omega = (B,\bm{b})$ is the outcome of a shadow shot with probability $p_\omega=\operatorname{Tr}(\rho E_\omega)$, with $E_\omega=\bigotimes_{i=1}^N[(\id+b_i\sigma_{B_i})/6]$.
The maximum likelihood estimator $
\hat\theta_{\mathrm{ML}}
=\arg\max_\theta\sum_{j=1}^n\log p_{\omega_j}$
attains the variance $1/(nF_\mathrm{cs})$, in the asymptotic shot limit. 

Moreover, $F_{\rm cs}$ is bounded by
\begin{equation}
    F_{\mathrm{cs}}[\rho]\le\frac{2NF_Q[\rho]}{2N+F_Q[\rho]}\le2N,
    \label{eq:shadow-information-hierarchy2}
\end{equation}
where $F_Q[\rho]$ is the quantum Fisher information (QFI) of the probe state $\rho$ for this encoding; the bound is saturated by suitably aligned pure product probes
(see the Supplemental Material~\cite{SM}). Combining \eqref{eq:hierarchyIK}, \eqref{eq:shadow-information-hierarchy1} and \eqref{eq:shadow-information-hierarchy2} we obtain the precision bound
\begin{equation}
 \begin{aligned}
  \Var_{\theta_0}(\thetahat_{\rm cs})&\ge
  \frac{1}{n}\left(\frac{1}{F_Q[\rho]}+\frac{1}{2N}\right).
 \end{aligned}
 \label{eq:shadow-precision-floor}
\end{equation}
The bound~\eqref{eq:shadow-precision-floor} separates the probe
contribution $1/F_Q[\rho]$ from the additional precision cost
$1/(2N)$ imposed by the randomized readout. In return, the shadow measurement provides access to arbitrary
observable expectation values, including those of an optimal
metrological observable, without having to identify or directly
measure that observable. The resulting estimator variance at fixed shot
count and small-error sample complexity at fixed accuracy and
confidence can scale no better than $N^{-1}$, even for nonclassical
probes with $F_Q[\rho]\propto N^2$. Nevertheless, as we show next,
entanglement can still provide constant-factor enhancements in
precision and corresponding reductions in measurement cost.

% The $1/2N$-term describes a noise penalty that the shadow protocol pays for extracting the huge number of considered observables, i.e., all Pauli strings up to weight $K$, from the same measurement data. The precision limit~(\ref{eq:shadow-precision-floor}) also tells us that using local Pauli shadows as the measurement ensemble cannot achieve variance and sample complexity scalings better than $O(N^{-1})$, even for nonclassical probes that achieve $F_Q[\rho]=O(N^2)$. However, as we will show next, there can still be improvements from entanglement in the form of constant-factor sensitivity enhancements.

\textit{Shadow-information bounds for separable states.---}We now discuss bounds on the shadow information for specific families of probe states. For any product state $\rho_{\mathrm{prod},\theta}=\bigotimes_{i=1}^{N}\rho_\theta^{(i)}$, we have
 \begin{equation}\label{eq:productbound}
     \mathcal{I}_K[\rho_{\mathrm{prod},\theta}] = \mathcal{I}_1[\rho_{\mathrm{prod},\theta}] \leq \frac{N}{3}.
 \end{equation}
 This bound is saturated whenever every local state is pure and polarized perpendicular to $\boldsymbol{n}$, such as a suitably aligned spin coherent state. 

\begin{figure*}[t]
  \centering
  \includegraphics[width=\textwidth]{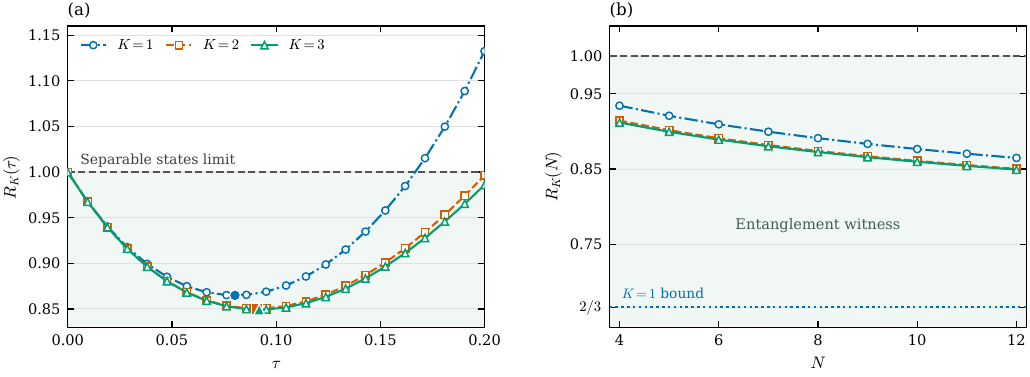}
  \caption{Entanglement-enhanced shadow precision with OAT probes.
  (a) Shadow information coefficient $R_K(\tau,N)=N/[3\mathcal{I}_K[\ket{\psi_\tau}]]$
  at $N=12$. Open markers show sampled twisting strengths; filled markers
  indicate minima. At fixed
  $\tau$, all points use the same probe and encoding.
  (b) $R_K(N)$, with $\tau$ optimized separately for each
  $(N,K)$ within $0\leq\tau\leq0.2$.
  The gray dashed line denotes the separable-state limit;
  the shaded region $R_K<1$ witnesses entanglement. The blue dotted line is
  the lower bound $R_1\geq2/3$, applicable only to $K=1$.
  Both panels use $\theta_0=-0.5$.}
  \label{fig:sample-complexity}
\end{figure*}
 
 To extend this result to all separable states $\rho_{\mathrm{sep},\theta}=\sum_\lambda p_\lambda\rho_{\lambda,\theta}^{(1)}\otimes\cdots\otimes \rho_{\lambda,\theta}^{(N)}$ with $p_{\lambda}\ge0$ and $\sum_{\lambda}p_{\lambda}=1$, we first establish that the shadow information is convex, i.e., for $\rho_{\mathrm{mix},\theta}=\sum_\lambda p_\lambda \rho_{\lambda,\theta}$, we have $\mathcal{I}[\rho_{\mathrm{mix},\theta}]
    \leq\sum_\lambda p_\lambda \mathcal{I}[\rho_{\lambda,\theta}]$ (see the Supplemental Material~\cite{SM}). Together with~(\ref{eq:productbound}), this implies that any separable probe is limited to
 \begin{equation}
    \mathcal{I}_{\mathrm{sep}}=\max_{\rho_{\mathrm{sep}}}\mathcal{I}_K[\rho_{\mathrm{sep},\theta}]=\frac{N}{3},
    \label{eq:shadow-info-convex}
 \end{equation}
 which is independent of $K$. Equation~\eqref{eq:shadow-info-convex} identifies the optimal
separable benchmark for the shadow-moment estimator.
Surpassing this benchmark, $\mathcal{I}_K[\rho]>\mathcal{I}_{\mathrm{sep}}$, certifies
entanglement through enhanced estimation precision.
The information gain relative to the ultimate precision benchmark~(\ref{eq:shadow-information-hierarchy2}), together with $F_Q[\rho]\le N^2$ for collective spin encoding~\cite{GiovannettiPRL2006,PezzePRL2009}, is bounded by
\begin{align}\label{eq:ultimatequantumbound}
    \mathcal{I}_K[\rho]\le \frac{2N^2}{N+2}<2N=6\mathcal{I}_{\mathrm{sep}},
\end{align}
i.e., limited by a factor of $6$ over the separable limit.

% Resolved: \mg{condense this section: frame as "naive" benchmark through direct measurements. move after discussion of shadow estimator properties.}
% Resolved: \lc{im not sure how should we reduce this?}
% \textit{Separate direct Pauli measurements.}---As a naive benchmark, consider measuring each of the $M=|\Pset|$ target Pauli strings separately, assigning $n/M$ independent shots to each. The moment estimates then have covariance $MV/n$, where $V_{PQ}=(1-\mu_P^2)\delta_{PQ}$, and the GLS estimator has variance
% \begin{equation}
%   \Var_{\theta_0}(\thetahat_{\rm dir})
%   =
%   \frac{M}{n}
%   \frac{1}{\boldsymbol{s}^T V^{-1}\boldsymbol{s}}.
%   \label{eq:direct-total-var}
% \end{equation}
% The explicit allocation factor $M$ arises from using separate shot pools, and is avoided by shadows which reuse the same randomized data across observables. This benchmark does not exploit observable grouping into commuting groups for optimized shot allocation; derivations, concentration bounds, and commuting group protocols are discussed in Ref.~\cite{SM}.

\textit{Numerical simulations.}---We illustrate how entanglement and nonlinear observables improve the protocol
using one-axis-twisted (OAT) probes,
$\ket{\psi_\tau}=\exp(-i\tau J_y^2)\ket{0}^{\otimes N}$, where
$J_\alpha=\frac{1}{2}\sum_i\sigma_\alpha^{(i)}$ and $\tau$ controls the
twisting strength~\cite{KitagawaPRA1993,PezzePRL2009,MolmerPRL1999}.
The parameter is encoded by a collective rotation
$\exp(-i\theta J_{\boldsymbol n})$, with $\boldsymbol n$ chosen along the
largest-variance collective-spin direction of each probe. For each $N$ and
cutoff $K=1,2,3$, we evaluate the local responses and shadow covariance
matrix for all Pauli strings of weight at most $K$ at $\theta_0=-0.5$,
and compute the corresponding shadow information $\mathcal I_K$.
We assume that the calibration and covariance matrix are known, and neglect
errors from estimating them with finite data. The calibration and optimization procedures are specified
in the Supplemental Material~\cite{SM}.

At $\tau=0$, the probe is a coherent spin state (CSS) perpendicular
to the encoding axis, saturating the separable benchmark
$\mathcal{I}_K=\mathcal{I}_{\mathrm{sep}}=N/3$ for every $K\ge1$.
Equation~\eqref{eq:shadow-var} then gives
$\Var_{\theta_0}(\thetahat_{\rm CSS})
=1/(n\mathcal{I}_{\mathrm{sep}})=3/(nN)$.
To quantify the improvement provided by OAT probes, we define the shadow information coefficient
\begin{equation}
 R_K(\tau,N)
 =\frac{\mathcal{I}_{\mathrm{sep}}}
 {\mathcal{I}_K[\ket{\psi_\tau}]}
 =\frac{N}{3\mathcal{I}_K[\ket{\psi_\tau}]}.
 \label{eq:normalized-shadow-cost}
\end{equation}
By Eq.~\eqref{eq:shadow-var}, this coefficient equals the variance
ratio
$R_K=\Var_{\theta_0}(\thetahat_{\rm cs})/
\Var_{\theta_0}(\thetahat_{\rm CSS})$
at fixed shot count and also compares the leading contributions
to the sufficient sample complexities in
Eq.~\eqref{eq:shadow-sample}.
Thus, $R_K=1$ for the CSS, while $R_K<1$ identifies
entanglement-enhanced precision relative to the optimal
separable benchmark.

Figure~\ref{fig:sample-complexity}(a) shows $R_K(\tau)$ at $N=12$.
Weak twisting lowers the ratio below one for all three cutoffs,
demonstrating an advantage from entanglement already at $K=1$.
Optimizing the twisting strength improves sensitivity by about $15\%$ compared
with the CSS, with most of the improvement already accessible through linear moments at $K=1$.

At each fixed $\tau$, the curves in Fig.~\ref{fig:sample-complexity}(a)
refer to the same probe and encoding. Their ordering therefore isolates the
benefit of including additional moments, consistently with the hierarchy~\eqref{eq:hierarchyIK}. Although individual higher-weight shadow
estimators carry larger noise factors, adding them cannot worsen the GLS
optimum. We also notice that for earlier twisting times, when the underlying state is approximately Gaussian, higher-weight Pauli strings, i.e., nonlinear observables, are not necessary to capture more information about the parameter. However, as we move into longer twisting times and, as a result, the underlying state becomes increasingly non-Gaussian, the addition of higher-weight Pauli strings becomes more relevant to lower the cost. 

Figure~\ref{fig:sample-complexity}(b) shows $R_K(N)$ after optimizing
$\tau$ separately for each $(N,K)$. Over the investigated range,
increasing $N$ strengthens the precision advantage relative to
the separable benchmark. The information bounds provide
quantitative reference values for this improvement:
Eq.~\eqref{eq:I1_bound} gives $R_1(N)\ge2/3$, while
Eq.~\eqref{eq:ultimatequantumbound} gives $R_K(N)\ge(N+2)/(6N)>1/6$ for every cutoff. Together, these results show that entanglement and additional moments
improve precision, but cannot overcome the scaling imposed by the randomized measurement. The Supplemental Material~\cite{SM} also compares separate
direct measurements and Twin-Fock probes.

\textit{Conclusion.---}We introduced a protocol for quantum parameter
estimation that combines randomized Pauli measurements with optimized
classical post-processing. The shadow information determines the
estimator variance and leading measurement cost from calibrated
responses and moment correlations. A separable benchmark
reveals entanglement-enhanced precision,
while higher-order moments can further improve precision for
non-Gaussian states.

The protocol is well suited to large noisy quantum circuits,
where optimal observables may be difficult to identify or access.
Since common decoherence mechanisms ultimately limit the probe
information to $F_Q[\rho]=O(N)$ at sufficiently large
$N$~\cite{HuelgaPRL1997,EscherNATUREPHYSICS2011,
DemkowiczNATURECOMMUNICATIONS2012}, randomized readout can offer
a powerful simplification at a modest additional precision cost:
the same measurement strategy applies to arbitrary probes and
encodings, without identifying or implementing the optimal
observable. This versatility makes
classical shadows a promising tool for robust quantum-enhanced
metrology under realistic conditions.

\section*{Acknowledgments}
This work was funded by the project PID2023-152724NA-I00, with funding from MCIU/AEI/10.13039/501100011033 and FSE+, by the project CNS2024-154818 with funding by MICIU/AEI /10.13039/501100011033, by the project CIPROM/2022/66 with funding by the Generalitat Valenciana, and by the CSIC Interdisciplinary Thematic Platform (PTI+) on Quantum Technologies (PTI-QTEP+). This work is supported through the project CEX2023-001292-S funded by MCIU/AEI. This work was supported by funding from Horizon Europe EU projects MSCA-SE CaLIGOLA, Project
ID: 101086123, and MSCA-DN CaLiForNIA, Project ID: 101119552.

\section*{Data availability statement}
Code that implements the proposed algorithm will be made available in the published version of this work. 

\bibliography{refs}

@FOOTNOTE{SM,
  note          = {See Supplemental Material for the estimator derivation, shadow covariance and information bounds, concentration bounds, direct and grouped measurement benchmarks, and numerical details.}
}

@article{GiovannettiSCIENCE2004,
  author        = {Giovannetti, Vittorio and
             Lloyd, Seth and
             Maccone, Lorenzo},
  title         = {Quantum-enhanced measurements: beating the standard quantum limit},
  journal       = {Science},
  volume        = {306},
  number        = {5700},
  pages         = {1330--1336},
  year          = {2004},
  doi           = {10.1126/science.1104149}
}

@article{GiovannettiPRL2006,
  author        = {Giovannetti, Vittorio and
             Lloyd, Seth and
             Maccone, Lorenzo},
  title         = {Quantum Metrology},
  journal       = {Phys. Rev. Lett.},
  volume        = {96},
  number        = {1},
  pages         = {010401},
  year          = {2006},
  doi           = {10.1103/PhysRevLett.96.010401}
}

@article{GiovannettiNATUREPHOTONICS2011,
  author        = {Giovannetti, Vittorio and
             Lloyd, Seth and
             Maccone, Lorenzo},
  title         = {Advances in quantum metrology},
  journal       = {Nat. Photonics},
  volume        = {5},
  pages         = {222--229},
  year          = {2011},
  doi           = {10.1038/nphoton.2011.35}
}

@article{PezzeRMP2018,
  author        = {Pezz{\`e}, Luca and
             Smerzi, Augusto and
             Oberthaler, Markus K. and
             Schmied, Roman and
             Treutlein, Philipp},
  title         = {Quantum metrology with nonclassical states of atomic ensembles},
  journal       = {Rev. Mod. Phys.},
  volume        = {90},
  number        = {3},
  pages         = {035005},
  year          = {2018},
  doi           = {10.1103/RevModPhys.90.035005}
}

@book{HelstromBOOK1976,
  author        = {Helstrom, Carl W.},
  title         = {Quantum Detection and Estimation Theory},
  publisher     = {Academic Press},
  address       = {New York},
  year          = {1976}
}

@book{HolevoBOOK1982,
  author        = {Holevo, Alexander S.},
  title         = {Probabilistic and Statistical Aspects of Quantum Theory},
  publisher     = {North-Holland},
  address       = {Amsterdam},
  year          = {1982}
}

@article{BraunsteinPRL1994,
  author        = {Braunstein, Samuel L. and
             Caves, Carlton M.},
  title         = {Statistical distance and the geometry of quantum states},
  journal       = {Phys. Rev. Lett.},
  volume        = {72},
  number        = {22},
  pages         = {3439--3443},
  year          = {1994},
  doi           = {10.1103/PhysRevLett.72.3439}
}

@article{WinelandPRA1992,
  author        = {Wineland, David J. and
             Bollinger, John J. and
             Itano, Wayne M. and
             Moore, Fred L. and
             Heinzen, Daniel J.},
  title         = {Spin squeezing and reduced quantum noise in spectroscopy},
  journal       = {Phys. Rev. A},
  volume        = {46},
  number        = {11},
  pages         = {R6797--R6800},
  year          = {1992},
  doi           = {10.1103/PhysRevA.46.R6797}
}

@article{KitagawaPRA1993,
  author        = {Kitagawa, Masahiro and
             Ueda, Masahito},
  title         = {Squeezed spin states},
  journal       = {Phys. Rev. A},
  volume        = {47},
  number        = {6},
  pages         = {5138--5143},
  year          = {1993},
  doi           = {10.1103/PhysRevA.47.5138}
}

@article{MaPHYSREP2011,
  author        = {Ma, Jian and
             Wang, Xiaoguang and
             Sun, C. P. and
             Nori, Franco},
  title         = {Quantum spin squeezing},
  journal       = {Phys. Rep.},
  volume        = {509},
  number        = {2--3},
  pages         = {89--165},
  year          = {2011},
  doi           = {10.1016/j.physrep.2011.08.003}
}

@article{PezzePRL2009,
  author        = {Pezz{\`e}, Luca and
             Smerzi, Augusto},
  title         = {Entanglement, nonlinear dynamics, and the {Heisenberg} limit},
  journal       = {Phys. Rev. Lett.},
  volume        = {102},
  number        = {10},
  pages         = {100401},
  year          = {2009},
  doi           = {10.1103/PhysRevLett.102.100401}
}

@article{GessnerPRL2019,
  author        = {Gessner, Manuel and
             Smerzi, Augusto and
             Pezz{\`e}, Luca},
  title         = {Metrological nonlinear squeezing parameter},
  journal       = {Phys. Rev. Lett.},
  volume        = {122},
  number        = {9},
  pages         = {090503},
  year          = {2019},
  doi           = {10.1103/PhysRevLett.122.090503}
}

@article{VolkoffFRONTIERS2024,
title = {Globally optimal interferometry with lossy twin {F}ock probes},
author = {Volkoff, T. J. and Ryu, Changhyun},
journal = {Frontiers in Physics},
volume = {12},
pages = {1369786},
year = {2024},
doi = {10.3389/fphy.2024.1369786}
}

@article{XuPRL2022,
  author        = {Xu, Kai and
             Zhang, Yu-Ran and
             Sun, Zheng-Hang and
             Li, Hekang and
             Song, Pengtao and
             Xiang, Zhongcheng and
             Huang, Kaixuan and
             Li, Hao and
             Shi, Yun-Hao and
             Chen, Chi-Tong and
             Song, Xiaohui and
             Zheng, Dongning and
             Nori, Franco and
             Wang, H. and
             Fan, Heng},
  title         = {Metrological characterization of non-{G}aussian entangled states of superconducting qubits},
  journal       = {Phys. Rev. Lett.},
  volume        = {128},
  number        = {15},
  pages         = {150501},
  year          = {2022},
  doi           = {10.1103/PhysRevLett.128.150501}
}

@article{HuangNATUREPHYSICS2020,
  author        = {Huang, Hsin-Yuan and
             Kueng, Richard and
             Preskill, John},
  title         = {Predicting many properties of a quantum system from very few measurements},
  journal       = {Nat. Phys.},
  volume        = {16},
  pages         = {1050--1057},
  year          = {2020},
  doi           = {10.1038/s41567-020-0932-7}
}

@article{ElbenPRL2020,
  author        = {Elben, Andreas and
             Kueng, Richard and
             Huang, Hsin-Yuan and
             van Bijnen, Rick and
             Kokail, Christian and
             Dalmonte, Marcello and
             Calabrese, Pasquale and
             Kraus, Barbara and
             Preskill, John and
             Zoller, Peter and
             Vermersch, Beno{\^i}t},
  title         = {Mixed-state entanglement from local randomized measurements},
  journal       = {Phys. Rev. Lett.},
  volume        = {125},
  number        = {20},
  pages         = {200501},
  year          = {2020},
  doi           = {10.1103/PhysRevLett.125.200501}
}

@article{StruchalinPRXQUANTUM2021,
  author        = {Struchalin, Grigorii I. and
             Zagorovskii, Ya. A. and
             Kovlakov, E. V. and
             Straupe, S. S. and
             Kulik, S. P.},
  title         = {Experimental Estimation of Quantum State Properties from Classical Shadows},
  journal       = {PRX Quantum},
  volume        = {2},
  number        = {1},
  pages         = {010307},
  year          = {2021},
  doi           = {10.1103/PRXQuantum.2.010307}
}

@article{AcharyaPRA2021,
  author        = {Acharya, Anurag and
             Saha, Sushant and
             Sengupta, Anirban M.},
  title         = {Shadow tomography based on informationally complete positive operator-valued measure},
  journal       = {Phys. Rev. A},
  volume        = {104},
  number        = {5},
  pages         = {052418},
  year          = {2021},
  doi           = {10.1103/PhysRevA.104.052418}
}

@article{ElbenNATUREREVIEWS2023,
  author        = {Elben, Andreas and
             Flammia, Steven T. and
             Huang, Hsin-Yuan and
             Kueng, Richard and
             Preskill, John and
             Vermersch, Beno{\^i}t and
             Zoller, Peter},
  title         = {The randomized measurement toolbox},
  journal       = {Nat. Rev. Phys.},
  volume        = {5},
  pages         = {9--24},
  year          = {2023},
  doi           = {10.1038/s42254-022-00535-2}
}

@article{HadfieldCMP2022,
  author        = {Hadfield, Charles and
             Bravyi, Sergey and
             Raymond, Rudy and
             Mezzacapo, Antonio},
  title         = {Measurements of quantum Hamiltonians with locally-biased classical shadows},
  journal       = {Commun. Math. Phys.},
  volume        = {391},
  pages         = {951--967},
  year          = {2022},
  doi           = {10.1007/s00220-022-04343-8}
}

@article{RathPRL2021,
  author        = {Rath, Aniket and
             Branciard, Cyril and
             Minguzzi, Anna and
             Vermersch, Beno{\^i}t},
  title         = {Quantum {Fisher} Information from Randomized Measurements},
  journal       = {Phys. Rev. Lett.},
  volume        = {127},
  number        = {26},
  pages         = {260501},
  year          = {2021},
  doi           = {10.1103/PhysRevLett.127.260501}
}

@article{MolmerPRL1999,
  author        = {M{\o}lmer, Klaus and
             S{\o}rensen, Anders},
  title         = {Multiparticle Entanglement of Hot Trapped Ions},
  journal       = {Phys. Rev. Lett.},
  volume        = {82},
  number        = {9},
  pages         = {1835--1838},
  year          = {1999},
  doi           = {10.1103/PhysRevLett.82.1835}
}

@book{BoucheronBOOK2013,
  author        = {Boucheron, St{\'e}phane and
             Lugosi, G{\'a}bor and
             Massart, Pascal},
  title         = {Concentration Inequalities: A Nonasymptotic Theory of Independence},
  publisher     = {Oxford University Press},
  address       = {Oxford},
  year          = {2013}
}

@article{Verteletskyi_2020,
  author        = {Verteletskyi, Vladyslav and
             Yen, Tzu-Ching and
             Izmaylov, Artur F.},
  title         = {Measurement optimization in the variational quantum eigensolver using a minimum clique cover},
  journal       = {J. Chem. Phys.},
  volume        = {152},
  number        = {12},
  pages         = {124114},
  year          = {2020},
  doi           = {10.1063/1.5141458}
}

@misc{hayashi2026finitesampleselectedcovariancespectra,
  author        = {Masahito Hayashi},
  title         = {Finite-Sample Selected Covariance Spectra in Classical Shadows},
  year          = {2026},
  eprint        = {2606.00527},
  archivePrefix = {arXiv},
  primaryClass  = {quant-ph}
}

@article{PhysRevA.99.032331,
  author        = {Schuld, Maria and
             Bergholm, Ville and
             Gogolin, Christian and
             Izaac, Josh and
             Killoran, Nathan},
  title         = {Evaluating analytic gradients on quantum hardware},
  journal       = {Phys. Rev. A},
  volume        = {99},
  pages         = {032331},
  year          = {2019},
  doi           = {10.1103/PhysRevA.99.032331}
}

@article{HuelgaPRL1997,
  author        = {Huelga, S. F. and
             Macchiavello, C. and
             Pellizzari, T. and
             Ekert, A. K. and
             Plenio, M. B. and
             Cirac, J. I.},
  title         = {Improvement of Frequency Standards with Quantum Entanglement},
  journal       = {Phys. Rev. Lett.},
  volume        = {79},
  number        = {20},
  pages         = {3865--3868},
  year          = {1997},
  doi           = {10.1103/PhysRevLett.79.3865}
}

@article{EscherNATUREPHYSICS2011,
  author        = {Escher, B. M. and
             de Matos Filho, R. L. and
             Davidovich, L.},
  title         = {General framework for estimating the ultimate precision limit in noisy quantum-enhanced metrology},
  journal       = {Nat. Phys.},
  volume        = {7},
  pages         = {406--411},
  year          = {2011},
  doi           = {10.1038/nphys1958}
}

@article{DemkowiczNATURECOMMUNICATIONS2012,
  author        = {Demkowicz-Dobrza{\'n}ski, R. and
             Ko{\l}ody{\'n}ski, J. and
             Gu{\c{t}}{\u{a}}, M.},
  title         = {The elusive {Heisenberg} limit in quantum-enhanced metrology},
  journal       = {Nat. Commun.},
  volume        = {3},
  pages         = {1063},
  year          = {2012},
  doi           = {10.1038/ncomms2067}
}

@article{ZhangPRL2025,
  author = {Zhang, Da-Jian and Tong, D. M.},
  title = {Krylov shadow tomography: Efficient estimation of quantum {Fisher} information},
  journal = {Phys. Rev. Lett.},
  volume = {134},
  pages = {110802},
  year = {2025},
  doi = {10.1103/PhysRevLett.134.110802}
}

@misc{MaoARXIV2026,
  author = {Mao, Liang and Chen, Senrui and Huang, Hsin-Yuan and Preskill, John and Zhou, Sisi},
  title = {Near-optimal quantum metrology with few-qubit measurements},
  year = {2026},
  eprint = {2608.01617},
  archivePrefix = {arXiv},
  primaryClass = {quant-ph}
}

@article{YuPRRESEARCH2021,
  author = {Yu, Min and Li, Dongxiao and Wang, Jingcheng and
            Chu, Yaoming and Yang, Pengcheng and Gong, Musang and
            Goldman, Nathan and Cai, Jianming},
  title = {Experimental estimation of the quantum {Fisher}
           information from randomized measurements},
  journal = {Phys. Rev. Research},
  volume = {3},
  pages = {043122},
  year = {2021},
  doi = {10.1103/PhysRevResearch.3.043122}
}

@article{ZhouPRXQUANTUM2026,
  author = {Zhou, Sisi and Chen, Senrui},
  title = {Randomized measurements for multiparameter quantum metrology},
  journal = {PRX Quantum},
  volume = {7},
  pages = {010314},
  year = {2026},
  doi = {10.1103/s27y-gbrp}
}

\clearpage
\onecolumngrid
\makeatletter
\onecolumn@grid@setup
\makeatother

\begin{center}
{\large\bfseries
Supplemental Material for\\
Quantum-enhanced metrology with classical shadows\par}

\vspace{1em}

{\normalsize
Luca Ion\textsuperscript{*} and Manuel Gessner\textsuperscript{$\dagger$}\par}

\vspace{0.5em}

{\small\itshape
Instituto de Física Corpuscular (IFIC),
CSIC-Universitat de València and
Departament de Física Teòrica, UV,
Av. Vicent Andrés Estellés, 19,
E-46100 Burjassot (València), Spain\par}
\end{center}

\begingroup
\renewcommand{\thefootnote}{\fnsymbol{footnote}}
\footnotetext[1]{
\href{mailto:luca.ion@uv.es}{luca.ion@uv.es}}

\footnotetext[2]{
\href{mailto:manuel.gessner@uv.es}{manuel.gessner@uv.es}}
\endgroup

\vspace{2em}

\appendix

\setcounter{secnumdepth}{3}

\renewcommand{\thesection}{\Alph{section}}
\renewcommand{\thesubsection}{\thesection.\arabic{subsection}}
\renewcommand{\thesubsubsection}{\thesubsection.\alph{subsubsection}}

\counterwithout*{equation}{section}

\setcounter{equation}{0}
\setcounter{figure}{0}
\setcounter{table}{0}

\renewcommand{\theequation}{S\arabic{equation}}
\renewcommand{\thefigure}{S\arabic{figure}}
\renewcommand{\thetable}{S\arabic{table}}

\section{Generalized least-squares estimator}
\label{app:gls}

This section derives the generalized least-squares (GLS) estimator used in the main text. We start from the calibrated local moment model
\begin{equation}
  \boldsymbol{\mu}(\theta)
  =
  \boldsymbol{\mu}(\theta_0)
  +
  \boldsymbol{s}(\theta-\theta_0)
  +
  O((\theta-\theta_0)^2),
\end{equation}
where $\boldsymbol{s}=\partial_\theta\boldsymbol{\mu}(\theta)|_{\theta_0}$. The measurement protocol produces an empirical moment vector $\boldsymbol{\muhat}$ with $\mathbb E_{\theta_0}[\boldsymbol{\muhat}]=\boldsymbol{\mu}(\theta_0)$. For $n$ independent shots, its covariance at the working point is written as \(\operatorname{Cov}_{\theta_0}(\boldsymbol{\muhat}) = \frac{1}{n}\Sigma,\) where $\Sigma$ is the corresponding single-shot covariance matrix.

We consider linear local estimators obtained from arbitrary linear combinations of the measured moments,
\begin{equation}
  \thetahat
  =
  \theta_0
  +
  \boldsymbol{a}^T
  \bigl[
    \boldsymbol{\muhat}
    -
    \boldsymbol{\mu}(\theta_0)
  \bigr].
  \label{eq:linear-estimator}
\end{equation}
Taking the expectation value and using the local expansion gives
\begin{equation}
  \mathbb E_\theta[\thetahat]
  =
  \theta_0
  +
  \boldsymbol{a}^T\boldsymbol{s}(\theta-\theta_0)
  +
  O((\theta-\theta_0)^2).
\end{equation}
Thus the estimator is locally unbiased at $\theta_0$ when $\boldsymbol{a}^T\boldsymbol{s}=1$. Its variance at the working point is \(\operatorname{Var}_{\theta_0}(\thetahat) = \frac{1}{n}\boldsymbol{a}^T\Sigma\boldsymbol{a}.\) The minimum-variance locally unbiased linear estimator is therefore obtained from the constrained problem
\begin{equation}
  \min_{\boldsymbol{a}}\,\boldsymbol{a}^T\Sigma\boldsymbol{a}
  \quad
  \text{subject to}
  \quad
  \boldsymbol{a}^T\boldsymbol{s}=1.
\end{equation}

Assume first that $\Sigma$ is nonsingular. Introducing a Lagrange multiplier, stationarity of \(f(\boldsymbol{a})=\boldsymbol{a}^T\Sigma\boldsymbol{a} - 2\lambda \bigl( \boldsymbol{a}^T\boldsymbol{s}-1 \bigr)\) gives $\Sigma\boldsymbol{a}=\lambda\boldsymbol{s}$. Imposing $\boldsymbol{a}^T\boldsymbol{s}=1$ yields
\begin{equation}
  \boldsymbol{a}_{\rm GLS}
  =
  \frac{\Sigma^{-1}\boldsymbol{s}}
       {\boldsymbol{s}^T\Sigma^{-1}\boldsymbol{s}} .
  \label{eq:a_gls}
\end{equation}
Substitution into the variance gives
\begin{equation}
  \operatorname{Var}_{\theta_0}(\thetahat_{\rm GLS})
  =
  \frac{1}{n}
  \frac{1}{\boldsymbol{s}^T\Sigma^{-1}\boldsymbol{s}} .
\end{equation}

The variance above is exact for every $n$ for this affine estimator with fixed, known calibration and covariance. The guarantees are local and assume fixed, known calibration and weights. At a differentiable interior working point, singular covariances are treated on their support, replacing inverses by Moore--Penrose pseudoinverses.

\section{Direct-measurement derivations}
\label{app:direct}

This appendix gives the details behind the separate direct-measurement baseline.  Each Pauli observable in \(\Pset\) is measured with its own independent shot pool, so the moment covariance is diagonal before the GLS combination is applied.

For each Pauli string \(P\in\Pset\), we perform \(R\) independent projective measurements of \(P\).  Let \(Z_{P,r}\in\{\pm1\}\) be the outcome of the \(r\)-th measurement.  The outcome has mean \(\E[Z_{P,r}]=\mu_P\), and since \(P^2=\id\), its single-shot variance is \(\Var(Z_{P,r})=1-\mu_P^2.\) The direct estimator of \(\mu_P\) is the sample average
\begin{equation}
  \widehat\mu^{\rm dir}_P=\frac{1}{R}\sum_{r=1}^{R}Z_{P,r}.
\end{equation}
This sample average is unbiased, with \(\E[\widehat\mu^{\rm dir}_P]=\mu_P\), and has variance \(\Var(\widehat\mu^{\rm dir}_P)=\frac{1-\mu_P^2}{R}.\) Stacking all components gives \(\boldsymbol{\muhat}^{\rm dir}=(\widehat\mu^{\rm dir}_P)_{P\in\Pset}.\) Since different Pauli strings are estimated from independent data, their covariances vanish: for \(P\neq Q\), \(\Cov(\widehat\mu^{\rm dir}_P,\widehat\mu^{\rm dir}_Q)=0.\) Consequently, \(\Cov(\boldsymbol{\muhat}^{\rm dir})=\frac{1}{R}V,\) where
\begin{equation}
  V_{PQ}=(1-\mu_P^2)\delta_{PQ}.
  \label{eq:direct-cov}
\end{equation}
This diagonal covariance is the essential structural difference from the shadow protocol: direct measurement does not reuse one experimental snapshot to estimate many Pauli observables.

Using the local model around \(\theta_0\), the direct linear estimator has the form
\begin{equation}
  \thetahat_{\rm dir}=\theta_0+\boldsymbol{a}^T(\boldsymbol{\muhat}^{\rm dir}-\boldsymbol{\mu}_0).
\end{equation}
Local unbiasedness imposes \(\boldsymbol{a}^T\boldsymbol{s}=1.\) Applying the GLS result with \(\Sigma=V\) gives
\begin{equation}
  \boldsymbol{a}_{\rm dir}=\frac{V^{-1}\boldsymbol{s}}{\boldsymbol{s}^TV^{-1}\boldsymbol{s}}.
\end{equation}
Hence
\begin{equation}
  \thetahat_{\rm dir}=\theta_0+\frac{\boldsymbol{s}^TV^{-1}(\boldsymbol{\muhat}^{\rm dir}-\boldsymbol{\mu}_0)}{\boldsymbol{s}^TV^{-1}\boldsymbol{s}}.
\end{equation}
The corresponding variance is
\begin{equation}
  \Var_{\theta_0}(\thetahat_{\rm dir})=\frac{1}{R}\frac{1}{\boldsymbol{s}^TV^{-1}\boldsymbol{s}}.
\end{equation}
Since \(V\) is diagonal,
\begin{equation}
  \boldsymbol{s}^TV^{-1}\boldsymbol{s}=\sum_{P\in\Pset}\frac{s_P^2}{1-\mu_P^2}.
  \label{appeq:dm_q_form}
\end{equation}
Equivalently,
\begin{equation}
  \Var_{\theta_0}(\thetahat_{\rm dir})=\frac{1}{R}\left(\sum_{P\in\Pset}\frac{s_P^2}{1-\mu_P^2}\right)^{-1}.
\end{equation}
The parameter \(R\) counts repetitions per Pauli.  Since \(M=\abs{\Pset}\), the total number of state preparations is \(n_{\rm dir}=RM.\) Thus
\begin{equation}
  \Var_{\theta_0}(\thetahat_{\rm dir})=\frac{M}{n_{\rm dir}}\frac{1}{\boldsymbol{s}^TV^{-1}\boldsymbol{s}}.
\end{equation}
In component form,
\begin{equation}
  \Var_{\theta_0}(\thetahat_{\rm dir})=\frac{M}{n_{\rm dir}}\left(\sum_{P\in\Pset}\frac{s_P^2}{1-\mu_P^2}\right)^{-1}.
\end{equation}

We next derive a concentration bound directly for the final scalar phase estimator.  Define
\begin{equation}
  Y^{\rm dir}_r=\sum_{P\in\Pset}a_{{\rm dir},P}(Z_{P,r}-\mu_P).
\end{equation}
Then
\begin{equation}
  \thetahat_{\rm dir}-\theta_0=\frac{1}{R}\sum_{r=1}^{R}Y^{\rm dir}_r.
\end{equation}
At the working point, \(\E[Y^{\rm dir}_r]=0.\) The independence of the direct Pauli measurement outcomes also gives \(\Var(Y^{\rm dir}_r)=\boldsymbol{a}_{\rm dir}^TV\boldsymbol{a}_{\rm dir}.\) For the GLS weights, this reduces to \(\Var(Y^{\rm dir}_r)=\frac{1}{\boldsymbol{s}^TV^{-1}\boldsymbol{s}}.\) Writing \(\sigma_{\rm dir}^2=\frac{1}{\boldsymbol{s}^TV^{-1}\boldsymbol{s}}\), we therefore obtain \(\Var(\thetahat_{\rm dir})=\frac{\sigma_{\rm dir}^2}{R}.\) For the boundedness term, \(Z_{P,r}\in\{\pm1\}\) and \(\abs{\mu_P}\le1\), so \(\abs{Z_{P,r}-\mu_P}\le2.\) Applying this bound to the weighted sum gives \(\abs{Y^{\rm dir}_r}\le2\norm{\boldsymbol{a}_{\rm dir}}_1.\) With the notation \(B_{\rm dir}=2\norm{\boldsymbol{a}_{\rm dir}}_1\), Bernstein's inequality gives
\begin{equation}
  \Pr\left(\abs{\thetahat_{\rm dir}-\theta_0}\ge\eta\right)\le2\exp\left[-\frac{R\eta^2}{2\sigma_{\rm dir}^2+\frac{2}{3}B_{\rm dir}\eta}\right].
\end{equation}
Equivalently, it is sufficient to choose
\begin{equation}
  R\ge\left(\frac{2\sigma_{\rm dir}^2}{\eta^2}+\frac{2B_{\rm dir}}{3\eta}\right)\log\frac{2}{\delta}
\end{equation}
to guarantee \(\abs{\thetahat_{\rm dir}-\theta_0}\le\eta\) with probability at least \(1-\delta\).  Since \(n_{\rm dir}=MR\),
\begin{equation}
  n_{\rm dir}\ge M\left(\frac{2\sigma_{\rm dir}^2}{\eta^2}+\frac{2B_{\rm dir}}{3\eta}\right)
\log\frac{2}{\delta}.
\end{equation}
In the small-error regime, the variance term gives the dominant scaling,
\begin{equation}
  n_{\rm dir}\sim\frac{M}{\eta^2}\frac{1}{\boldsymbol{s}^TV^{-1}\boldsymbol{s}}\log\frac{1}{\delta}.
  \label{eq:direct-sample}
\end{equation}

Finally, we derive a conservative sufficient condition for the separate direct-measurement benchmark. Using \(1-\mu_P^2\le1,\) we have \(\frac{s_P^2}{1-\mu_P^2}\ge s_P^2.\) Summing these inequalities and using Eq.~\eqref{appeq:dm_q_form} gives
\begin{equation}
  \frac{1}{\boldsymbol{s}^TV^{-1}\boldsymbol{s}}\le\frac{1}{\norm{\boldsymbol{s}}_2^2}.
\end{equation}
The looser but more uniform form used for comparison is
\begin{equation}
  \frac{1}{\norm{\boldsymbol{s}}_2^2}\le\frac{\norm{\boldsymbol{s}}_1^2}{\norm{\boldsymbol{s}}_2^4}.
\end{equation}
Substituting this bound into the variance-dominated sample complexity gives
\begin{equation}
  n_{\rm dir}\gtrsim\frac{M}{\eta^2}\frac{\norm{\boldsymbol{s}}_1^2}{\norm{\boldsymbol{s}}_2^4}\log\frac{1}{\delta}.
  \label{eq:suff-dm}
\end{equation}
The variance-dominated expression requires the small  error regime  $B_{\rm dir}\eta\ll3\sigma_{\rm dir}^2$.

\section{Local Pauli shadow covariance}
\label{app:shadow-covariance}

This appendix derives the single-shot covariance matrix for local Pauli shadows.  Let \(\Gamma_{PQ}=\Cov(X_P,X_Q),\) where \(X_P\) denotes the single-shot shadow estimator of the Pauli string \(P\).

For local Pauli shadows,
\begin{equation}
  X_P=3^{w_P}\mathbf{1}[B\sim P]\prod_{j\in\supp(P)}b_j.
\end{equation}
Here \(w_P=\wt(P)\), the random product basis is \(B\in\{X,Y,Z\}^{\otimes N}\), and \(b_j\in\{\pm1\}\) is the outcome on qubit \(j\).  The event \(B\sim P\) means that the sampled basis hits every non-identity site of \(P\): \(B_j=P_j\) for all \(j\in\supp(P)\).  This estimator is unbiased, so that \(\E[X_P]=\mu_P.\)

The covariance depends on whether two Pauli strings are qubit-wise compatible. If $P \not\sim Q$, then there is a site \(j\) such that both strings are non-identity but require different local bases: \(P_j\neq\id,\) \(Q_j\neq\id\), and \(P_j\neq Q_j.\) No product basis can hit both strings on that site.  Hence \(X_PX_Q=0\) on every shadow shot, and \(\E[X_PX_Q]=0\), which gives \(\Gamma_{PQ}=-\mu_P\mu_Q.\)

If \(P\) and \(Q\) are qubit-wise compatible, define \(I_{PQ}=\supp(P)\cap\supp(Q).\) On the intersection, the two strings have the same local Pauli.  Therefore the duplicated local outcomes square to one in the product \(X_PX_Q\), and the remaining product corresponds to \(PQ\).  The resulting second moment is \(\E[X_PX_Q]=3^{\abs{I_{PQ}}}\langle PQ\rangle\), so that the covariance is \(\Gamma_{PQ}=3^{\abs{I_{PQ}}}\langle PQ\rangle-\mu_P\mu_Q.\) Combining the compatible and incompatible cases gives
\begin{equation}
  \Gamma_{PQ}=\begin{cases}3^{\abs{I_{PQ}}}\langle PQ\rangle-\mu_P\mu_Q, & P\sim Q, \\ -\mu_P\mu_Q, & P\not\sim Q.\end{cases}
  \label{eq:sm-shadow-covariance}
\end{equation}
For \(P=Q\), this reduces to \(\Gamma_{PP}=3^{w_P}-\mu_P^2.\) Since the shadow snapshots are independent, the empirical covariance is \(\Cov(\boldsymbol{\muhat})=\frac{1}{n}\Gamma.\) Its diagonal entries therefore give \(\Var(\widehat\mu_P)=\frac{3^{w_P}-\mu_P^2}{n}.\) 
\section{Shadow-information hierarchy and precision bounds}
\label{app:shadow-information-bounds}

We prove Eqs.~(15) and (16) of the main text
for independent, uniformly random local Pauli measurements and the encoding
$\rho_\theta=e^{-i\theta H}\rho_0 e^{i\theta H}$ with
$H=\frac12\sum_i\boldsymbol n\cdot\boldsymbol\sigma_i$ and
$\norm{\boldsymbol n}=1$. All quantities below are evaluated at the same
working point $\theta_0$, and we write $\rho$ for $\rho_{\theta_0}$.

\subsection{POVM reconstruction and added noise}

Each measurement record $\omega=(B,\boldsymbol b)$ specifies the independently and uniformly sampled local Pauli bases $B_i\in\{x,y,z\}$ and their outcomes $b_i\in\{-1,+1\}$. For a fixed basis string $B$, the Born rule gives
\begin{equation}
 p(\boldsymbol b\mid B)=\Tr\!\left[\rho\bigotimes_{i=1}^N\Pi_{B_i,b_i}\right],
 \qquad \Pi_{B_i,b_i}=\frac{\id+b_i\sigma_{B_i}}{2}.
\end{equation}
Since $p(B)=3^{-N}$, the joint record probability is
\begin{equation}
 p_\omega=p(B)p(\boldsymbol b\mid B)=\Tr(\rho E_\omega),
 \qquad E_\omega=3^{-N}\bigotimes_{i=1}^N\Pi_{B_i,b_i}
 =\bigotimes_{i=1}^N\frac{\id+b_i\sigma_{B_i}}{6}.
\end{equation}
These effects are positive and satisfy $\sum_\omega E_\omega=\id$: for each fixed $B$, summing the outcome projectors over $\boldsymbol b$ gives the identity, and $\sum_Bp(B)=1$. Thus the complete measurement record is described by a positive-operator-valued measure (POVM). This construction does not assume independent measurement outcomes.

The corresponding classical-shadow snapshot is obtained by applying the inverse local measurement channel to each outcome projector,
\begin{equation}
 \hat\rho(\omega)=\bigotimes_{i=1}^N(3\Pi_{B_i,b_i}-\id).
\end{equation}
It is a random Hermitian estimator, not necessarily a positive density operator, and obeys $\E[\hat\rho]=\rho$. Hence, for any Hermitian observable $A$, the canonical shadow statistic $X_A(\omega)=\Tr[A\hat\rho(\omega)]$ satisfies $\E[X_A]=\Tr(\rho A)$. Because this holds for every input state, it implies the operator identity
\begin{equation}
 \sum_\omega X_A(\omega)E_\omega=A.
 \label{eq:bounds-operator-unbiasedness}
\end{equation}
For the noise argument, let $f$ be any real statistic reconstructing $A$, so that $\sum_\omega f(\omega)E_\omega=A$; the canonical choice is $f=X_A$. Define
\begin{equation}
 \nu_f=\Var(f)-\Var_\rho(A)
 =\sum_\omega p_\omega f(\omega)^2-\langle A^2\rangle.
\end{equation}
Expanding the positive operator
\begin{align}
 R_f&=\sum_\omega(f\id-A)E_\omega(f\id-A)\nonumber\\
 &=\sum_\omega f^2E_\omega-A^2\succeq0
\end{align}
using the reconstruction identity for $f$ gives
$\nu_f=\Tr(\rho R_f)\ge0$.

Use the canonical shadow statistic $X_H$ to reconstruct $H$ on this same outcome space. Introduce
\begin{equation}
 U_\omega=\sqrt{E_\omega}(f\id-A)\sqrt\rho,\qquad
 W_\omega=\sqrt{E_\omega}(X_H\id-H)\sqrt\rho.
\end{equation}
Their squared direct-sum Hilbert--Schmidt norms are $\nu_f$ and $\nu_H:=\Var(X_H)-\Var_\rho(H)$.
The cross term therefore satisfies
\begin{equation}
 T_{fH}:=\sum_\omega\Tr(U_\omega^\dagger W_\omega)
 =\E[fX_H]-\langle AH\rangle,\qquad
 |T_{fH}|^2\le\nu_f\nu_H.
\end{equation}
The classical product is real, whereas
$\langle HA\rangle=\langle AH\rangle^*$. Consequently,
\begin{equation}
 s_A=\langle i[H,A]\rangle=-2\operatorname{Im}T_{fH},\qquad
 s_A^2\le4\nu_f\nu_H.
 \label{eq:bounds-noise-uncertainty}
\end{equation}
Now to compute $\nu_H$, write $H=\sum_i h_i$ with
$h_i=\boldsymbol n\cdot\boldsymbol\sigma_i/2$. Its canonical estimator is
$X_H=\sum_i X_{h_i}$, where $X_{h_i}=3n_{B_i}b_i/2$. Uniform basis sampling
and $b_i^2=1$ imply \(\E[X_{h_i}^2]=\frac34,\qquad h_i^2=\frac14\id.\) For distinct sites, independent basis sampling and local reconstruction give
\begin{equation}
 \E[X_{h_i}X_{h_j}]
 =\frac14\sum_{\alpha,\beta}n_\alpha n_\beta
 \langle\sigma_\alpha^{(i)}\sigma_\beta^{(j)}\rangle
 =\langle h_i h_j\rangle.
\end{equation}
This identity does not assume independent signs or an uncorrelated probe.
All cross-site contributions cancel in the added variance, leaving
\begin{equation}
 \nu_H=\E[X_H^2]-\langle H^2\rangle=\frac N2.
\end{equation}
Substitution into Eq.~\eqref{eq:bounds-noise-uncertainty} proves
\begin{equation}
 \nu_f\ge\frac{s_A^2}{2N}.
 \label{eq:bounds-added-noise}
\end{equation}
In particular, taking $f=X_A$ gives $\nu_A:=\Var(X_A)-\Var_\rho(A)\ge s_A^2/(2N)$ for every linear combination of the Pauli observables at any cutoff $K$.
Expanding \eqref{eq:bounds-added-noise} gives
the following bound, where the subscript $A$ denotes an arbitrary observable with nonzero response:
\begin{equation}
 \Var(X_A)\ge
 \left(\chi_A^2+\frac1{2N}\right)s_A^2,
 \quad \chi_A^2 = \frac{\Var_\rho(A)}{s_A^2}.
 \label{eq:bounds-statistic-variance}
\end{equation}
The method-of-moments sensitivity for measuring $A$ obeys the standard QFI bound~\cite{GessnerPRL2019},
\begin{equation}
    \chi_A^{-2} \leq F_Q.
    \label{eq:qfi-bound-nls}
\end{equation}
For the canonical statistic of $A=\sum_{P\in\Pset}a_PP$, one has $\Var(X_A)=\boldsymbol a^T\Gamma_K\boldsymbol a$ and $s_A=\boldsymbol a^T\boldsymbol s_K$. Choose the shadow-GLS weights $\boldsymbol a_*=\Gamma_K^+\boldsymbol s_K/\mathcal I_K$, assuming $\mathcal I_K>0$, and define $A_*=\sum_Pa_{*,P}P$. Then $s_{A_*}=1$, and Eq.~\eqref{eq:bounds-statistic-variance} yields
\begin{equation}
 {1\over\mathcal I_K}\ge\chi^2[\rho]+{1\over2N},\qquad
 \mathcal I_K\le{2N\chi^{-2}[\rho]\over2N+\chi^{-2}[\rho]},\qquad
 \chi^{-2}[\rho]={s_{A_*}^2\over\Var_\rho(A_*)}.
 \label{eq:app-gls-squeezing-bound}
\end{equation}
Here $\chi^{-2}[\rho]=\chi_{A_*}^{-2}$: it is the sensitivity of the observable selected by shadow GLS, with the cutoff and encoding held fixed.

To distinguish it from a different optimization, define
\begin{equation}
 \chi_{\rm ideal,opt}^2=
 \min_{\boldsymbol a^T\boldsymbol s_K=1}
 \Var_\rho\!\left(\sum_Pa_PP\right).
\end{equation}
The ideal observable covariance has entries $\langle\{P,Q\}\rangle/2-\mu_P\mu_Q$, whereas shadow GLS minimizes the quadratic form of $\Gamma_K$. Therefore $\chi_{\rm ideal,opt}^2\le\chi^2[\rho]$, generally with different optimizers. The bound with $\chi_{\rm ideal,opt}^2$ is also valid, but weaker than evaluating the added-noise inequality at $A_*$. These two quantities must not be identified~\cite{GessnerPRL2019}.

Combining Eqs.~\eqref{eq:bounds-statistic-variance} and \eqref{eq:qfi-bound-nls} gives $\Var(X_A)\ge s_A^2[1/F_Q+1/(2N)]$ for nonzero QFI.

For $\mathcal I_K>0$, define the optimal single-shot variance coefficient $\sigma_K^2:=1/\mathcal I_K$, so that $\Var_{\theta_0}(\thetahat_{\rm cs})=\sigma_K^2/n$.
For nested moment sets, extend any feasible level-$K$ coefficient vector to level $K+1$ by zero padding. Its slope and shadow variance are unchanged. Hence $\sigma_{K+1}^2\le\sigma_K^2$, or $\mathcal I_{K+1}\ge\mathcal I_K$. This comparison fixes the state, encoding, working point, measurement ensemble, and calibration; it does not require assigning nonzero weight to the newly included moments.
\subsection{Classical Fisher information bound}

For any fixed real statistic $f(\omega)$, define
$A_f=\sum_\omega f(\omega)E_\omega$. Then $f$ reconstructs $A_f$ as an
operator identity, so the general added-noise inequality~\eqref{eq:bounds-added-noise}, together with the QFI bound~\eqref{eq:qfi-bound-nls}, gives $\Var(f)\ge s_{A_f}^2[1/F_Q+1/(2N)]$, where $s_{A_f}=\partial_\theta\E_\theta[f]|_{\theta_0}$. The statistic $f$ need not equal the canonical shadow statistic $X_{A_f}$. At a regular
working point, take the score frozen at $\theta_0$,
\begin{equation}
 q(\omega)=\left.\partial_\theta\log p_\omega(\theta)\right|_{\theta_0},
 \qquad F_{\rm cs}=\sum_{\omega:p_\omega>0}
 \frac{(\partial_\theta p_\omega)^2}{p_\omega}.
\end{equation}
Its mean is zero, its variance is $F_{\rm cs}$, and the derivative of its
mean, holding the statistic fixed, is also $F_{\rm cs}$. Thus
\begin{equation}
 F_{\rm cs}\ge\left(\frac1{F_Q}+\frac1{2N}\right)F_{\rm cs}^2,
 \qquad F_{\rm cs}\le\frac{2NF_Q}{2N+F_Q}\le2N.
\end{equation}

For any fixed statistic $f$, its response is its covariance with the score:
\begin{equation}
 \partial_\theta\E_\theta[f]\big|_{\theta_0}
 =\E[(f-\E[f])q],\qquad
 |\partial_\theta\E_\theta[f]|^2\le\Var(f)F_{\rm cs}.
\end{equation}
Applying this to $f=X_A=\sum_{P\in\Pset}a_PX_P$ and maximizing its response-to-variance ratio proves $\mathcal I_K\le F_{\rm cs}$. Equality requires the centered statistic to be proportional to the score on its probability support. Having all Pauli moments available does not ensure this condition.

\subsection{Stronger bound for weight-one sector}

For $A=\sum_i\boldsymbol a_i\cdot\boldsymbol\sigma_i$, let
$\boldsymbol r_i$ denote the local Bloch vectors. The same cross-site noise
cancellation as above gives
\begin{equation}
 \nu_A=2\sum_i\norm{\boldsymbol a_i}^2,\qquad
 s_A=\sum_i\boldsymbol a_i\cdot(\boldsymbol n\times\boldsymbol r_i).
\end{equation}
Since $\norm{\boldsymbol r_i}\le1$, Cauchy--Schwarz implies
$s_A^2\le N\sum_i\norm{\boldsymbol a_i}^2$. Therefore
\begin{equation}
 \Var(X_A)\ge\nu_A\ge\frac{2s_A^2}{N},\qquad
 \sigma_1^2\ge\frac2N,\qquad \mathcal I_1\le\frac N2.
\end{equation}
This stronger constant is specific to the weight-one space and does not
replace the all-$K$ ceiling.

\subsection{Explicit saturation and maximum-likelihood efficiency}

Consider the collective rotation
\begin{equation}
 \rho_\theta=e^{-i\theta J_y}\rho_0e^{i\theta J_y},\qquad
 J_y=\frac12\sum_{i=1}^N\sigma_y^{(i)},
\end{equation}
acting on a product of identical pure qubits,
\begin{equation}
 \rho_0=\bigl(\lvert\psi_\phi\rangle\langle\psi_\phi\rvert\bigr)^{\otimes N},\qquad
 \lvert\psi_\phi\rangle=\cos(\phi/2)\lvert0\rangle+\sin(\phi/2)\lvert1\rangle.
\end{equation}
Here $\phi$ is a fixed, known preparation angle and $\theta$ is the unknown parameter. We evaluate the information at the working point $\theta_0=0$.

First specialize to $N=1$. The encoded Bloch vector is $\boldsymbol r(\theta)=(\sin(\phi+\theta),0,\cos(\phi+\theta))$, so at $\theta_0=0$ one has $\boldsymbol r=(\sin\phi,0,\cos\phi)$ and $\dot{\boldsymbol r}=(\cos\phi,0,-\sin\phi)$. The six local record probabilities and their derivatives are
\begin{equation}
 p_{\alpha,b}={1+b r_\alpha\over6},\qquad
 \dot p_{\alpha,b}={b\dot r_\alpha\over6},\qquad \alpha=x,y,z,\quad b=\pm1.
\end{equation}
For $|r_\alpha|<1$, the Fisher-information contribution from basis $\alpha$, including its sampling probability, is
\begin{equation}
 \sum_{b=\pm1}{\dot p_{\alpha,b}^2\over p_{\alpha,b}}
 ={\dot r_\alpha^2\over3(1-r_\alpha^2)}.
\end{equation}
Choose $\sin\phi\cos\phi\ne0$. The $x$ and $z$ contributions are each $1/3$ and the $y$ contribution vanishes, giving $F_{\rm cs}^{(1)}=2/3$. Meanwhile $F_Q^{(1)}=4\Var(\sigma_y/2)=1$.

Returning to the $N$-qubit product probe, the record distribution factorizes. The total score is a sum of independent zero-mean local scores, so its variance, and hence its Fisher information, adds:
\begin{equation}
 F_Q=N,\qquad F_{\rm cs}={2N\over3}
 ={2NF_Q\over2N+F_Q}.
 \label{eq:app-fisher-saturation}
\end{equation}
The \hyperref[app:css-benchmark]{product-state appendix} gives instead $\mathcal I_K=N/3$ for every $K\ge1$. Thus this example saturates the Fisher-information bound, not the moment-estimator inequality $\mathcal I_K\le F_{\rm cs}$. It does not demonstrate attainability of the universal $2N$ ceiling with entangled probes.

Under the usual regularity and local-identifiability assumptions, maximum likelihood is asymptotically efficient, with variance $1/(nF_{\rm cs})+o(n^{-1})$. This statement applies locally at regular working points.

\section{Shadow concentration and sufficient condition}
\label{app:concentration}

This appendix derives the concentration bound for the shadow-GLS phase estimator and the conservative sufficient condition quoted in the main text.

Let
\begin{equation}
  \boldsymbol{a}_{\rm GLS}=\frac{\Gamma^{-1}\boldsymbol{s}}{\boldsymbol{s}^T\Gamma^{-1}\boldsymbol{s}}.
\end{equation}
For the \(\ell\)-th shadow snapshot, we collect the single-shot estimators into \(\boldsymbol{X}^{(\ell)}=(X_P^{(\ell)})_{P\in\Pset}.\) The empirical moment vector is then \(\boldsymbol{\muhat}=\frac{1}{n}\sum_{\ell=1}^{n}\boldsymbol{X}^{(\ell)}.\) The phase-estimator error can be written as the empirical average
\begin{equation}
  \thetahat_{\rm GLS}-\theta_0=\frac{1}{n}\sum_{\ell=1}^{n}Y_\ell,
\end{equation}
where \(Y_\ell=\boldsymbol{a}_{\rm GLS}^T(\boldsymbol{X}^{(\ell)}-\boldsymbol{\mu}_0).\) At the working point, this centered variable has mean \(\E[Y_\ell]=0\) and single-shot variance \(\Var(Y_\ell)=\boldsymbol{a}_{\rm GLS}^T\Gamma \boldsymbol{a}_{\rm GLS}.\) The GLS identity then yields \(\Var(Y_\ell)=\frac{1}{\boldsymbol{s}^T\Gamma^{-1}\boldsymbol{s}}.\) Writing \(\sigma_{\rm GLS}^2=\frac{1}{\boldsymbol{s}^T\Gamma^{-1}\boldsymbol{s}}\), the variance of the empirical average becomes \(\Var(\thetahat_{\rm GLS})=\frac{\sigma_{\rm GLS}^2}{n}.\)

For Bernstein's inequality we also need to bound $|Y_\ell|$.  Since \(\abs{X_P^{(\ell)}}\le3^{w_P}\) and \(\abs{\mu_P}\le1\), we have \(\abs{X_P^{(\ell)}-\mu_P}\le3^{w_P}+1.\) Applying this inequality to the weighted sum gives \(\abs{Y_\ell}\le\sum_{P\in\Pset}\abs{a_{{\rm GLS},P}}(3^{w_P}+1).\) Since every target Pauli has weight at most \(K\), this implies \(\abs{Y_\ell}\le(3^K+1)\norm{\boldsymbol{a}_{\rm GLS}}_1.\) Retaining the individual Pauli weights gives the tighter range bound
\begin{equation}
 R_{\rm GLS}=\sum_{P\in\Pset}|a_{{\rm GLS},P}|(3^{w_P}+1)
 \le(3^K+1)\norm{\boldsymbol a_{\rm GLS}}_1.
 \label{eq:app-shadow-range}
\end{equation}
Bernstein's inequality gives
\begin{equation}
  \Pr\left(\abs{\thetahat_{\rm GLS}-\theta_0}\ge\eta\right)\le2\exp\left[-\frac{n\eta^2}{2\sigma_{\rm GLS}^2+\frac{2}{3}R_{\rm GLS}\eta}\right].
\end{equation}
Equivalently, it is sufficient to take
\begin{equation}
  n\ge\left(\frac{2\sigma_{\rm GLS}^2}{\eta^2}+\frac{2R_{\rm GLS}}{3\eta}\right)\log\frac{2}{\delta},
  \label{eq:app-shadow-finite-budget}
\end{equation}
to obtain \(\abs{\thetahat_{\rm GLS}-\theta_0}\le\eta\) with probability at least \(1-\delta\).  In the small-error regime the variance term dominates, giving
\begin{equation}
  n\sim\frac{1}{\eta^2}\frac{1}{\boldsymbol{s}^T\Gamma^{-1}\boldsymbol{s}}\log\frac{1}{\delta}.
\end{equation}

We now derive a simpler sufficient condition by upper bounding \((\boldsymbol{s}^T\Gamma^{-1}\boldsymbol{s})^{-1}\).  Since GLS minimizes \(\boldsymbol{a}^T\Gamma\boldsymbol{a}\) under \(\boldsymbol{a}^T\boldsymbol{s}=1\), any feasible trial vector gives
\begin{equation}
  \frac{1}{\boldsymbol{s}^T\Gamma^{-1}\boldsymbol{s}}=\min_{\boldsymbol{a}:\boldsymbol{a}^T\boldsymbol{s}=1}\boldsymbol{a}^T\Gamma\boldsymbol{a}\le \boldsymbol{a}^T\Gamma\boldsymbol{a}.
\end{equation}
The useful trial vector for local Pauli shadows is \(b_P=\frac{s_P3^{-w_P}}{A_s},\) with normalization \(A_s=\sum_{P\in\Pset}s_P^2 3^{-w_P}.\) It is locally unbiased because
\begin{equation}
  \boldsymbol{b}^T\boldsymbol{s}=\frac{1}{A_s}\sum_{P\in\Pset}s_P^2 3^{-w_P}=1.
\end{equation}
Therefore
\begin{equation}
  \frac{1}{\boldsymbol{s}^T\Gamma^{-1}\boldsymbol{s}}\le \boldsymbol{b}^T\Gamma \boldsymbol{b}.
\end{equation}
To bound this quadratic form, we introduce the second-moment matrix \(C_{PQ}=\E[X_PX_Q].\) Since \(\Gamma=C-\boldsymbol{\mu}\boldsymbol{\mu}^T\) and \(\boldsymbol{\mu}\boldsymbol{\mu}^T\succeq0\), we have \(\Gamma\preceq C\), and hence \(\boldsymbol{b}^T\Gamma \boldsymbol{b}\le \boldsymbol{b}^TC\boldsymbol{b}.\) Substituting the trial vector gives
\begin{equation}
  \boldsymbol{b}^TC\boldsymbol{b}=\frac{1}{A_s^2}\sum_{P,Q}s_Ps_Q3^{-w_P}3^{-w_Q}C_{PQ}.
\end{equation}
Taking absolute values,
\begin{equation}
  \boldsymbol{b}^TC\boldsymbol{b}\le\frac{1}{A_s^2}\sum_{P,Q}\abs{s_P}\abs{s_Q}3^{-w_P}3^{-w_Q}\abs{C_{PQ}}.
\end{equation}
For local Pauli shadows, \(C_{PQ}=3^{\abs{I_{PQ}}}\langle PQ\rangle\) when \(P\sim Q\), while \(C_{PQ}=0\) when \(P\not\sim Q\).  Here the overlap of the two supports is denoted by \(I_{PQ}=\supp(P)\cap\supp(Q).\) Both cases are therefore covered by the bound \(\abs{C_{PQ}}\le\mathbf{1}[P\sim Q]3^{\abs{I_{PQ}}}.\) It follows that
\begin{equation}
  \boldsymbol{b}^TC\boldsymbol{b}\le\frac{1}{A_s^2}\sum_{P\sim Q}\abs{s_P}\abs{s_Q}3^{\abs{I_{PQ}}-w_P-w_Q}.
\end{equation}
Using \(w_P+w_Q-\abs{I_{PQ}}=\abs{\supp(P)\cup\supp(Q)},\) we obtain
\begin{equation}
  \frac{1}{\boldsymbol{s}^T\Gamma^{-1}\boldsymbol{s}}\le\frac{1}{A_s^2}\sum_{P\sim Q}\abs{s_P}\abs{s_Q}3^{-\abs{\supp(P)\cup\supp(Q)}}.
\end{equation}
This expression already captures the shadow-specific support structure.  For the simpler bound in the main text, we use \(3^{-\abs{\supp(P)\cup\supp(Q)}}\le1\) and drop the compatibility restriction.  Then
\begin{equation}
  \sum_{P\sim Q}\abs{s_P}\abs{s_Q}3^{-\abs{\supp(P)\cup\supp(Q)}}\le\norm{\boldsymbol{s}}_1^2.
\end{equation}
Therefore
\begin{equation}
  \frac{1}{\boldsymbol{s}^T\Gamma^{-1}\boldsymbol{s}}\le\frac{\norm{\boldsymbol{s}}_1^2}{A_s^2}.
\end{equation}
Since \(w_P\le K\), the normalization satisfies \(A_s\ge3^{-K}\norm{\boldsymbol{s}}_2^2.\) Squaring this bound gives \(A_s^2\ge3^{-2K}\norm{\boldsymbol{s}}_2^4.\) Combining the last two inequalities gives
\begin{equation}
  \frac{1}{\boldsymbol{s}^T\Gamma^{-1}\boldsymbol{s}}\le3^{2K}\frac{\norm{\boldsymbol{s}}_1^2}{\norm{\boldsymbol{s}}_2^4}.
\end{equation}
Substitution into the variance-dominated concentration scaling yields
\begin{equation}
  n\gtrsim\frac{3^{2K}}{\eta^2}\frac{\norm{\boldsymbol{s}}_1^2}{\norm{\boldsymbol{s}}_2^4}\log\frac{1}{\delta}.
\end{equation}

The variance-dominated expression applies in the small error regime $R_{\rm GLS}\eta\ll3\sigma_{\rm GLS}^2=3/\mathcal I_K$.

\section{Grouped direct-measurement baseline}
\label{app:grouped}

This appendix describes the grouped direct-measurement baseline mentioned in the main text.  The target Pauli set is partitioned into \(G\) commuting groups, where each Pauli is contained in exactly one group: \(\Pset=\bigsqcup_{g=1}^{G}C_g.\) Within each group, the observables satisfy \([P,Q]=0\) for all \(P,Q\in C_g\).  We use only the abstract fact that each commuting family admits a simultaneous eigenbasis; circuit costs for implementing that basis are not included in the shot-complexity accounting.

Grouping can reduce the allocation overhead relative to measuring every Pauli separately. Finding a partition into the minimum number of commuting groups is a minimum clique-cover problem on the graph whose edges connect commuting Pauli strings, and is NP-hard in general~\cite{Verteletskyi_2020}. In the next appendix section we give a simple, nonoptimal local-basis construction that reduces the leading-order number of settings from ${N\choose K}3^K$ to ${N\choose K}2^K$ at fixed $K$.

For each group \(C_g\), we perform \(R\) repeated joint measurements in a basis diagonalizing all \(P\in C_g\). A single shot \(r\) of group \(g\) produces simultaneous eigenvalue outcomes \(Z_P^{(g,r)}\in\{\pm1\}\) for all \(P\in C_g\). The corresponding group estimator is
\begin{equation}
  \widehat\mu^{\rm grp}_P=\frac{1}{R}\sum_{r=1}^{R}Z_P^{(g(P),r)}.
\end{equation}
This estimator is unbiased, with \(\E_\theta[\widehat\mu^{\rm grp}_P]=\mu_P(\theta).\)

The covariance is block diagonal in the group decomposition.  If \(P,Q\in C_g\), the same joint shot estimates both observables and \(Z_P^{(g,r)}Z_Q^{(g,r)}=Z_{PQ}^{(g,r)}.\) Taking the expectation gives \(\E_\theta[Z_PZ_Q]=\langle PQ\rangle_{\rho_\theta}.\) The single-shot grouped covariance is therefore \(\Omega_{PQ}(\theta)=\langle PQ\rangle_{\rho_\theta}-\mu_P(\theta)\mu_Q(\theta)\) when \(P\) and \(Q\) belong to the same group, and \(\Omega_{PQ}(\theta)=0\) when they belong to different groups.  Therefore, after ordering Pauli strings group by group, we can write \(\Omega(\theta)=\bigoplus_{g=1}^{G}\Omega^{(g)}(\theta).\) Averaging over the repeated measurements gives the empirical covariance \(\Cov_\theta(\boldsymbol{\muhat}^{\rm grp})=\frac{1}{R}\Omega(\theta).\)

Around \(\theta_0\), a locally unbiased grouped estimator has the form
\begin{equation}
  \thetahat_{\rm grp}=\theta_0+\boldsymbol{a}^T(\boldsymbol{\muhat}^{\rm grp}-\boldsymbol{\mu}_0),
\end{equation}
with \(\boldsymbol{a}^T\boldsymbol{s}=1.\) The variance of this estimator is \(\Var(\thetahat_{\rm grp})=\frac{1}{R}\boldsymbol{a}^T\Omega \boldsymbol{a}.\) GLS gives
\begin{equation}
  \boldsymbol{a}_{\rm grp}=\frac{\Omega^{-1}\boldsymbol{s}}{\boldsymbol{s}^T\Omega^{-1}\boldsymbol{s}}.
\end{equation}
Thus
\begin{equation}
  \thetahat_{\rm grp}=\theta_0+\frac{\boldsymbol{s}^T\Omega^{-1}(\boldsymbol{\muhat}^{\rm grp}-\boldsymbol{\mu}_0)}{\boldsymbol{s}^T\Omega^{-1}\boldsymbol{s}}.
\end{equation}
The variance becomes
\begin{equation}
  \Var(\thetahat_{\rm grp})=\frac{1}{R}\frac{1}{\boldsymbol{s}^T\Omega^{-1}\boldsymbol{s}}.
\end{equation}
Since the protocol uses \(R\) shots for each of the \(G\) groups, the total shot count is \(n_{\rm grp}=GR.\) Hence
\begin{equation}
  \Var_{\theta_0}(\thetahat_{\rm grp})=\frac{G}{n_{\rm grp}}\frac{1}{\boldsymbol{s}^T\Omega^{-1}\boldsymbol{s}}.
\end{equation}

The grouped estimator also admits a scalar Bernstein bound.  For this purpose, we write \(\sigma_{\rm grp}^2=\frac{1}{\boldsymbol{s}^T\Omega^{-1}\boldsymbol{s}}.\) We also write the GLS weight vector according to the group decomposition as \(\boldsymbol{a}_{\rm grp}=(\boldsymbol{a}^{(1)},\ldots,\boldsymbol{a}^{(G)}),\) and set \(B_{\rm grp}=2\max_g\norm{\boldsymbol{a}^{(g)}}_1.\) Since \(\abs{Z_P-\mu_P}\le2\), each group-level scalar summand is bounded by \(B_{\rm grp}\).  Bernstein's inequality gives
\begin{equation}
  \Pr\left(\abs{\thetahat_{\rm grp}-\theta_0}\ge\eta\right)\le2\exp\left[-\frac{n_{\rm grp}\eta^2}{G\left(2\sigma_{\rm grp}^2+\frac{2}{3}B_{\rm grp}\eta\right)}\right].
\end{equation}
Equivalently, it is sufficient to take
\begin{equation}
  n_{\rm grp}\gtrsim G\left(\frac{2\sigma_{\rm grp}^2}{\eta^2}+\frac{2B_{\rm grp}}{3\eta}\right)\log\frac{2}{\delta}.
\end{equation}
In the variance-dominated regime,
\begin{equation}
  n_{\rm grp}\sim\frac{G}{\eta^2}\frac{1}{\boldsymbol{s}^T\Omega^{-1}\boldsymbol{s}}\log\frac{1}{\delta}.
\end{equation}

We finish with the grouped sufficient condition.  By the GLS variational identity,
\begin{equation}
  \frac{1}{\boldsymbol{s}^T\Omega^{-1}\boldsymbol{s}}=\min_{\boldsymbol{a}:\boldsymbol{a}^T\boldsymbol{s}=1}\boldsymbol{a}^T\Omega \boldsymbol{a}.
\end{equation}
Using the feasible trial vector \(\boldsymbol{a}=\frac{\boldsymbol{s}}{\norm{\boldsymbol{s}}_2^2},\) we obtain
\begin{equation}
  \frac{1}{\boldsymbol{s}^T\Omega^{-1}\boldsymbol{s}}\le\frac{\boldsymbol{s}^T\Omega \boldsymbol{s}}{\norm{\boldsymbol{s}}_2^4}.
\end{equation}
Because \(\Omega\) is block diagonal,
\begin{equation}
  \boldsymbol{s}^T\Omega \boldsymbol{s}=\sum_{g=1}^{G}{\boldsymbol{s}^{(g)}}^T\Omega^{(g)}\boldsymbol{s}^{(g)}.
\end{equation}
For a fixed group,
\begin{equation}
  {\boldsymbol{s}^{(g)}}^T\Omega^{(g)}\boldsymbol{s}^{(g)}=\Var\left(\sum_{P\in C_g}s_PZ_P\right).
\end{equation}
Using \(\Var(X)\le\E[X^2]\),
\begin{equation}
  {\boldsymbol{s}^{(g)}}^T\Omega^{(g)}\boldsymbol{s}^{(g)}\le\sum_{P,Q\in C_g}\abs{s_P}\abs{s_Q}\abs{\langle PQ\rangle_{\rho_{\theta_0}}}.
\end{equation}
Since \(\abs{\langle PQ\rangle_{\rho_{\theta_0}}}\le1\), each group obeys the bound \({\boldsymbol{s}^{(g)}}^T\Omega^{(g)}\boldsymbol{s}^{(g)}\le\norm{\boldsymbol{s}^{(g)}}_1^2.\) Combining the group bounds gives
\begin{equation}
  \frac{1}{\boldsymbol{s}^T\Omega^{-1}\boldsymbol{s}}\le\frac{\sum_{g=1}^{G}\norm{\boldsymbol{s}^{(g)}}_1^2}{\norm{\boldsymbol{s}}_2^4}.
\end{equation}
Finally,
\begin{equation}
  \sum_{g=1}^{G}\norm{\boldsymbol{s}^{(g)}}_1^2\le\left(\sum_{g=1}^{G}\norm{\boldsymbol{s}^{(g)}}_1\right)^2=\norm{\boldsymbol{s}}_1^2.
\end{equation}
Therefore
\begin{equation}
  \frac{1}{\boldsymbol{s}^T\Omega^{-1}\boldsymbol{s}}\le\frac{\norm{\boldsymbol{s}}_1^2}{\norm{\boldsymbol{s}}_2^4}.
\end{equation}
Substituting this into the variance-dominated grouped sample complexity yields
\begin{equation}
  n_{\rm grp}\gtrsim\frac{G}{\eta^2}\frac{\norm{\boldsymbol{s}}_1^2}{\norm{\boldsymbol{s}}_2^4}\log\frac{1}{\delta}.
\end{equation}

\section{Simple commuting-group construction}
\label{app:commuting-construction}

This appendix gives the explicit local-basis grouping construction. The construction is deliberately simple: it groups Pauli strings according to a compatible local Pauli basis.  It is not intended to be optimal.

For a Pauli string \(P\in\Pset\), we define the associated full local basis string \(B(P)=B_1(P)\otimes\cdots\otimes B_N(P),\) where
\begin{equation}
  B_i(P)=
  \begin{cases}
    P_i, & P_i\neq \id,\\
    Z, & P_i=\id.
  \end{cases}
\end{equation}
Thus identity factors of \(P\) are completed with local \(Z\) measurements.  For each possible basis string \(B\), we collect the compatible strings into \(C_B=\{P\in\Pset: B(P)=B\}.\) These sets form the partition \(\Pset=\bigsqcup_B C_B\), whose union is disjoint because every Pauli string has a unique completed basis \(B(P)\).

Each set \(C_B\) is qubit-wise commuting.  Indeed, if \(P,Q\in C_B\), then on every site either both nonidentity factors coincide with the same local basis element, or at least one of the factors is the identity.  Hence \(P\) and \(Q\) commute locally on every qubit, and therefore \([P,Q]=0.\) All observables in \(C_B\) can therefore be measured in the same product basis \(B\).  This gives a valid grouped direct-measurement protocol.

It remains to count the number of groups.  A completed basis string \(B\) can contain local \(X\) or \(Y\) entries only on sites where the original Pauli string was nonidentity.  Therefore, if \(B\) contains \(t\) non-\(Z\) entries, then \(t\le K\).  For fixed \(t\), one chooses the \(t\) sites in \({N\choose t}\) ways and chooses either \(X\) or \(Y\) on each of them, giving \(2^t\) choices.  Hence this construction uses
\begin{equation}
  G=\sum_{t=0}^{K}{N\choose t}2^t
  \label{eq:qb-group-count}
\end{equation}
qubit-wise commuting groups.  By contrast, measuring every Pauli string separately uses
\begin{equation}
  M=\abs{\Pset}=\sum_{w=1}^{K}{N\choose w}3^w
  \label{eq:separate-count}
\end{equation}
settings.  At fixed \(K\) and large \(N\), the leading terms scale as \({N\choose K}2^K\) and \({N\choose K}3^K\), respectively.

\section{Product-state shadow information bound}
\label{app:css-benchmark}

Here we establish a shadow-GLS information bound for arbitrary product
states, proving Eq.~(17) of the main text and its
saturation by a coherent spin state (CSS), in three steps. We first
calculate the shadow-GLS information carried by the weight-one Pauli
observables of a single qubit. We then combine the $N$ independent qubits of
the product state. Finally, we show that higher-weight Pauli strings contain no
additional local GLS information for this product-state family.
We then extend the same upper bound to all separable states by convexity
of the shadow-GLS information.

\subsection{Single-qubit case}

Consider an arbitrary product probe, whose local states need not be identical
or pure. Under collective phase encoding generated by
$J_{\boldsymbol n}=\frac12\sum_{i=1}^{N}\boldsymbol n\cdot\boldsymbol\sigma^{(i)}$,
with $\norm{\boldsymbol n}=1$, it remains a product state,
\begin{equation}
 \rho_\theta=\bigotimes_{i=1}^{N}\rho_\theta^{(i)}.
 \label{eq:css-product-state}
\end{equation}
For the single-qubit calculation, consider qubit $1$ and write its Bloch
vector as $\boldsymbol{r}(\theta)$, so that
\begin{equation}
 \rho_\theta^{(1)}
 =\frac{1}{2}\left(I_2+\sum_{\alpha=x,y,z}r_\alpha(\theta)\sigma_\alpha\right),
 \qquad
 r_\alpha(\theta)=\Tr\!\left(\rho_\theta^{(1)}\sigma_\alpha\right).
\end{equation}
We denote its response at the working point by
$\dot{\boldsymbol{r}}=\partial_\theta\boldsymbol{r}|_{\theta_0}$. If
$\boldsymbol{n}$ is the unit encoding axis, the unitary rotation of the Bloch
vector gives
\begin{equation}
 \dot{\boldsymbol{r}}=\boldsymbol{n}\times\boldsymbol{r}.
 \label{eq:css-bloch-rotation}
\end{equation}
Since $\norm{\boldsymbol{r}}\leq1$, we have
\begin{equation}
 \boldsymbol{r}^{T}\dot{\boldsymbol{r}}=0,
 \qquad
 \norm{\dot{\boldsymbol{r}}}^{2}
 =\norm{\boldsymbol{r}}^{2}-(\boldsymbol{n}^{T}\boldsymbol{r})^{2}\leq1.
 \label{eq:css-bloch-response}
\end{equation}
The first relation follows for any unitary qubit rotation; equality in the
second holds precisely for a pure local state polarized perpendicular to
the encoding axis.

We now calculate the covariance of the three single-qubit Pauli shadow
estimators, with the local measurement basis chosen uniformly from $x,y,z$.
A simple application of Eq.~\eqref{eq:sm-shadow-covariance} of the main text gives
\begin{equation}
 \Gamma^{(1)}_{\alpha\beta}
 =3\delta_{\alpha\beta}-r_\alpha r_\beta,
 \qquad
 \Gamma^{(1)}=3I_3-\boldsymbol{r}\boldsymbol{r}^{T}.
 \label{eq:css-single-qubit-covariance}
\end{equation}
Using $\boldsymbol{r}^{T}\dot{\boldsymbol{r}}=0$, we obtain \(\Gamma^{(1)}\dot{\boldsymbol{r}} =3\dot{\boldsymbol{r}}.\) Thus, when nonzero, $\dot{\boldsymbol{r}}$ is an eigenvector of $\Gamma^{(1)}$ with
eigenvalue $3$. Multiplying the preceding equation by
$(\Gamma^{(1)})^{-1}$ gives
\begin{equation}
 \left(\Gamma^{(1)}\right)^{-1}\dot{\boldsymbol{r}}
 =\frac{1}{3}\dot{\boldsymbol{r}}.
 \label{eq:css-inverse-on-response}
\end{equation}
The shadow-GLS information contributed by one qubit is consequently
\begin{equation}
 \dot{\boldsymbol{r}}^{T}\left(\Gamma^{(1)}\right)^{-1}
 \dot{\boldsymbol{r}}
 =\frac{1}{3}\norm{\dot{\boldsymbol{r}}}^{2}\leq\frac{1}{3}.
 \label{eq:css-one-qubit-information}
\end{equation}

\subsection{Combining multiple qubits}

Write $\boldsymbol r_i$ and $\dot{\boldsymbol r}_i$ for the Bloch vector and
response of qubit $i$, evaluated at the working point, and set
$\Gamma^{(i)}=3I_3-\boldsymbol r_i\boldsymbol r_i^T$.
Let $X_\alpha^{(i)}$ denote the weight-one shadow estimator of
$\sigma_\alpha^{(i)}$. For two distinct qubits $i\neq j$, the product state
in Eq.~\eqref{eq:css-product-state} and the independent local basis choices
give
\begin{align}
 \E\!\left[X_\alpha^{(i)}X_\beta^{(j)}\right]
 &=\left\langle\sigma_\alpha^{(i)}\sigma_\beta^{(j)}\right\rangle
 =r_{i,\alpha}r_{j,\beta},\nonumber\\
 \E\!\left[X_\alpha^{(i)}\right]
 \E\!\left[X_\beta^{(j)}\right]
 &=r_{i,\alpha}r_{j,\beta}.
\end{align}
Hence
\begin{equation}
 \Cov\!\left(X_\alpha^{(i)},X_\beta^{(j)}\right)=0,
 \qquad i\neq j.
 \label{eq:css-cross-qubit-zero}
\end{equation}
Although all qubits are measured in the same global shot, their weight-one
shadow fluctuations are therefore uncorrelated for this product state.

Let $L$ denote the complete weight-one sector. Its covariance and response
have the block forms
\begin{equation}
 \Gamma_{LL}
 =\bigoplus_{i=1}^{N}\Gamma^{(i)},
 \qquad
 \boldsymbol{s}_{L}
 =\begin{pmatrix}
   \dot{\boldsymbol{r}}_1\\[-1mm]
   \vdots\\[-1mm]
   \dot{\boldsymbol{r}}_N
  \end{pmatrix}.
 \label{eq:css-weight-one-blocks}
\end{equation}
The inverse of a block-diagonal matrix is obtained by inverting each block.
Using Eq.~\eqref{eq:css-one-qubit-information}, the total weight-one GLS
information is therefore
\begin{align}
 \boldsymbol{s}_{L}^{T}\Gamma_{LL}^{-1}\boldsymbol{s}_{L}
 &=\sum_{i=1}^{N}
   \dot{\boldsymbol{r}}_i^{T}\left(\Gamma^{(i)}\right)^{-1}
   \dot{\boldsymbol{r}}_i\nonumber\\
 &=\frac{1}{3}\sum_{i=1}^{N}\norm{\dot{\boldsymbol r}_i}^{2}
 \leq\frac{N}{3}.
 \label{eq:css-weight-one-information}
\end{align}

\subsection{Why higher-weight moments add no information}

Before treating a general Pauli string, consider the two-body example
$P=\sigma_\alpha^{(i)}\sigma_\gamma^{(j)}$. Product-state factorization and
the product rule give
\begin{equation}
 \mu_P=r_{i,\alpha} r_{j,\gamma},
 \qquad
 s_P=\dot r_{i,\alpha} r_{j,\gamma}+r_{i,\alpha}\dot r_{j,\gamma}.
 \label{eq:css-two-body-response}
\end{equation}
The two terms are precisely the responses inherited from qubits $i$ and $j$.
This relation also appears in the covariance with the weight-one sector. For
example, the covariance with a Pauli observable on qubit $i$ is \(\Gamma_{P,(i,\beta)} =3\delta_{\beta,\alpha}r_{j,\gamma}-r_{i,\alpha} r_{j,\gamma} r_{i,\beta}.\) Contracting this covariance with the response of qubit $i$ gives
\begin{align}
 \frac{1}{3}\sum_{\beta=x,y,z}
 \Gamma_{P,(i,\beta)}\dot r_{i,\beta}
 &=\dot r_{i,\alpha} r_{j,\gamma}
 -\frac{r_{i,\alpha} r_{j,\gamma}}{3}
   \boldsymbol{r}_i^{T}\dot{\boldsymbol{r}}_i\nonumber\\
 &=\dot r_{i,\alpha} r_{j,\gamma}.
\end{align}
The analogous contraction for qubit $j$ gives
$r_{i,\alpha}\dot r_{j,\gamma}$. Their sum is exactly the full response $s_P$ in
Eq.~\eqref{eq:css-two-body-response}.

The same identity holds at arbitrary weight. Let
$P=\bigotimes_{i\in S}\sigma_{p_i}^{(i)}$ have support $S$. Then
\begin{align}
 \mu_P&=\prod_{i\in S}r_{i,p_i},\nonumber\\
 s_P&=\sum_{j\in S}\dot r_{j,p_j}
       \prod_{i\in S\setminus\{j\}}r_{i,p_i}.
 \label{eq:css-string-response}
\end{align}
For $j\notin S$, the covariance of $P$ with the weight-one estimator on
qubit $j$ vanishes. For $j\in S$, the local-Pauli-shadow covariance is
\begin{equation}
 \Gamma_{P,(j,\beta)}
 =3\delta_{\beta,p_j}\prod_{i\in S\setminus\{j\}}r_{i,p_i}
  -\mu_P r_{j,\beta}.
 \label{eq:css-cross-covariance}
\end{equation}
Equations~\eqref{eq:css-inverse-on-response} and
\eqref{eq:css-cross-covariance} imply
\begin{align}
 \sum_{j=1}^{N}\sum_{\beta=x,y,z}
 \Gamma_{P,(j,\beta)}
 \left(\Gamma_{LL}^{-1}\boldsymbol{s}_{L}\right)_{(j,\beta)}
 &=\frac{1}{3}\sum_{j\in S}\sum_{\beta=x,y,z}
   \Gamma_{P,(j,\beta)}\dot r_{j,\beta}\nonumber\\
 &=s_P.
 \label{eq:css-string-redundancy-expanded}
\end{align}
Collecting this equality for every higher-weight string gives the block
identity
\begin{equation}
 \boldsymbol{s}_{H}
 =\Gamma_{HL}\Gamma_{LL}^{-1}\boldsymbol{s}_{L},
 \label{eq:css-redundant-response}
\end{equation}
where $H$ denotes all included strings of weight at least two. In words, the
response of the higher-weight sector is completely reproduced by its
covariance with the weight-one response. It has no residual response after
the weight-one sector has been accounted for.

To connect this identity to GLS, write the full covariance in blocks,
\begin{equation}
 \Gamma=
 \begin{pmatrix}
  \Gamma_{LL} & \Gamma_{LH}\\
  \Gamma_{HL} & \Gamma_{HH}
 \end{pmatrix}.
\end{equation}
The block-inverse identity gives
\begin{align}
 \boldsymbol{s}^{T}\Gamma^{-1}\boldsymbol{s}
 ={}&\boldsymbol{s}_{L}^{T}\Gamma_{LL}^{-1}\boldsymbol{s}_{L}\nonumber\\
 &+\boldsymbol{s}_{H|L}^{T}S_H^{-1}\boldsymbol{s}_{H|L},
 \label{eq:css-schur-information}
\end{align}
where
\begin{equation}
 \boldsymbol{s}_{H|L}
 =\boldsymbol{s}_{H}-\Gamma_{HL}\Gamma_{LL}^{-1}\boldsymbol{s}_{L},
 \qquad
 S_H=\Gamma_{HH}-\Gamma_{HL}\Gamma_{LL}^{-1}\Gamma_{LH}.
\end{equation}
Here $\boldsymbol{s}_{H|L}$ is the response left in the higher-weight sector
after its correlations with the weight-one sector are removed, and $S_H$ is
the corresponding residual covariance. Equation~\eqref{eq:css-redundant-response}
shows that $\boldsymbol{s}_{H|L}=0$. The second term in
Eq.~\eqref{eq:css-schur-information} therefore vanishes, and higher-weight
moments cannot increase the GLS information. The same conclusion holds on
the covariance support, with Moore--Penrose pseudoinverses, if a block is
singular.

Combining this result with Eq.~\eqref{eq:css-weight-one-information} yields,
for every cutoff $K\geq1$,
\begin{equation}
 \boldsymbol{s}^{T}\Gamma^{-1}\boldsymbol{s}
 =\frac{1}{3}\sum_{i=1}^{N}
 \left[\norm{\boldsymbol r_i}^{2}-(\boldsymbol n^T\boldsymbol r_i)^2\right]
 \leq\frac{N}{3}.
 \label{eq:css-exact-variance-coefficient}
\end{equation}
The bound is saturated whenever every local state is pure and polarized
perpendicular to $\boldsymbol n$; identical local polarizations are not
required. In particular, at $\tau=0$ a pure CSS with its encoding axis
orthogonal to its polarization saturates the bound. For this CSS, the
leading sample complexity therefore reduces to
\begin{equation}
 \widetilde n_{\rm CSS}(N)
 =\frac{3\log(1/\delta)}{\eta^2N}.
\end{equation}

\subsection{Extension to separable states}
\label{app:separable-shadow-benchmark}

Consider a separable probe under the same collective encoding,
\begin{equation}
 \rho_\theta=\sum_\lambda p_\lambda\rho_{\lambda,\theta},
 \qquad
 \rho_{\lambda,\theta}=\bigotimes_{i=1}^{N}\rho_{\lambda,\theta}^{(i)},
 \qquad p_\lambda\geq0,\quad\sum_\lambda p_\lambda=1,
\end{equation}
where the weights are independent of $\theta$. Fix $K$ and the uniform
local-Pauli measurement and reconstruction protocol. Let
$\boldsymbol X=(X_P)_{P\in\Pset}$ be the vector of single-shot
estimators. The index $\lambda$ labels a component state, not a snapshot:
$\boldsymbol\mu_\lambda$, $\boldsymbol s_\lambda$ and $\Gamma_\lambda$
are its estimator mean, response and shadow covariance at the working point.
All expectations below are classical averages over random bases and outcomes.

For the complete record $\omega=(B,\boldsymbol b)$, with $\Pi_{B,\boldsymbol b}=\bigotimes_i(\id+b_i\sigma_{B_i})/2$ and $\Pr(B)=3^{-N}$, the Born rule gives
\begin{equation}
 \Pr(\omega\mid\rho_\theta)
 =3^{-N}\Tr(\rho_\theta\Pi_{B,\boldsymbol b})
 =\sum_\lambda p_\lambda\Pr(\omega\mid\rho_{\lambda,\theta}).
 \label{eq:sep-shadow-record-mixture}
\end{equation}
Since $\boldsymbol X(\omega)$ is the same function of the record for every
component, both its first and second moments are linear under mixing:
\begin{equation}
 \boldsymbol\mu=\sum_\lambda p_\lambda\boldsymbol\mu_\lambda,
 \qquad
 \E_\rho[\boldsymbol X\boldsymbol X^T]
 =\sum_\lambda p_\lambda
   \E_{\rho_\lambda}[\boldsymbol X\boldsymbol X^T].
\end{equation}
Differentiating the mean gives
$\boldsymbol s=\sum_\lambda p_\lambda\boldsymbol s_\lambda$.
The classical law of total covariance therefore yields
\begin{align}
 \Gamma(\rho)
 &=\sum_\lambda p_\lambda\Gamma_\lambda
   +\sum_\lambda p_\lambda
    (\boldsymbol\mu_\lambda-\boldsymbol\mu)
    (\boldsymbol\mu_\lambda-\boldsymbol\mu)^T\nonumber\\
 &\succeq\sum_\lambda p_\lambda\Gamma_\lambda.
 \label{eq:sep-shadow-covariance-concavity}
\end{align}
The extra term is positive semidefinite because its quadratic form is
$\sum_\lambda p_\lambda
[\boldsymbol a^T(\boldsymbol\mu_\lambda-\boldsymbol\mu)]^2\geq0$.

The information admits the following representation:
\begin{equation}
 \mathcal{I}_K[\rho]:=\boldsymbol s^T\Gamma(\rho)^{-1}\boldsymbol s
 =\max_{\boldsymbol a\neq0}
   \frac{(\boldsymbol a^T\boldsymbol s)^2}
        {\boldsymbol a^T\Gamma(\rho)\boldsymbol a}.
\end{equation}
Set $t_\lambda=\boldsymbol a^T\boldsymbol s_\lambda$ and
$v_\lambda=\boldsymbol a^T\Gamma_\lambda\boldsymbol a$ for a fixed
$\boldsymbol a$. Covariance concavity and weighted Cauchy--Schwarz give
\begin{equation}
 \frac{(\boldsymbol a^T\boldsymbol s)^2}
      {\boldsymbol a^T\Gamma(\rho)\boldsymbol a}
 \leq\frac{(\sum_\lambda p_\lambda t_\lambda)^2}
            {\sum_\lambda p_\lambda v_\lambda}
 \leq\sum_\lambda p_\lambda\frac{t_\lambda^2}{v_\lambda}
 \leq\sum_\lambda p_\lambda \mathcal{I}_K[\rho_\lambda].
\end{equation}
Maximizing over $\boldsymbol a$ proves convexity of $\mathcal{I}_K$.
Applying the product-state bound~\eqref{eq:css-exact-variance-coefficient}
to every component then gives
\begin{equation}
 \mathcal{I}_K[\rho_{\rm sep}]
 \leq\sum_\lambda p_\lambda \mathcal{I}_K[\rho_\lambda]
 \leq\frac{N}{3},\qquad K\geq1.
 \label{eq:sep-shadow-information-bound}
\end{equation}
The bound is tight, since a pure CSS polarized perpendicular to the encoding
axis saturates it. The equality $\mathcal{I}_K=\mathcal{I}_1$ need not survive mixing: higher-weight
moments may improve the information for a separable state, while still obeying
Eq.~\eqref{eq:sep-shadow-information-bound}.

\section{Numerical implementation}
\label{app:numerics}

Here we summarize the procedure used to obtain Fig.~1 of the main text and the additional comparisons. The simulations evaluate covariance-aware GLS variance coefficients from state-vector expectation values; they do not emulate finite-shot measurement records. Numerical differentiation, interpolation, and matrix inversion introduce deterministic numerical error, distinct from finite-shot calibration error, which is excluded. The implementation, parameter files, and plotting routines are contained in the accompanying codebase.

\subsection{Probe preparation and parameter encoding}

The OAT probe is prepared as
\begin{equation}
 \ket{\psi_\tau}=e^{-i\tau J_y^2}\ket{0}^{\otimes N},
 \qquad J_\alpha=\frac{1}{2}\sum_{i=1}^N\sigma_\alpha^{(i)}.
\end{equation}
In the circuit implementation, the twisting operation is decomposed into all-to-all two-qubit gates $\exp[-i\tau Y_iY_j/2]$; this is equivalent to $\exp(-i\tau J_y^2)$ up to a global phase.  The phase is subsequently encoded through $\exp(-i\theta J_{\boldsymbol n})$.  For every $(N,\tau)$, the direction $\boldsymbol n$ is chosen as the eigenvector associated with the largest eigenvalue of the $3\times3$ collective-spin covariance matrix.  Thus, the encoding direction maximizes $\Var(J_{\boldsymbol n})$ within the family of collective rotations. The encoding axis is fixed before phase encoding and shared across all moment cutoffs; it maximizes the probe QFI.

For fixed $N$ and cutoff $K$, we generate the complete set
\begin{equation}
 \Pset=\{P:1\leq\wt(P)\leq K\},
 \qquad |\Pset|=\sum_{k=1}^K {N\choose k}3^k.
\end{equation}
For each $P\in\Pset$, its expectation value $\mu_P$ and local response $s_P$ are evaluated at $\theta_0=-0.5$.  The response is obtained from a 401-point calibration curve over $[-\pi/2+0.01,0]$: a centered finite difference is applied on the grid and the resulting derivative is linearly interpolated to $\theta_0$.  All observables are retained in the calculations reported here, including those with vanishing first-order response, since correlations with responsive observables can still affect the GLS solution.

\subsection{Covariances and sample complexity}

The direct covariance $V$ and shadow covariance $\Gamma$ are assembled using Eqs.~\eqref{eq:direct-cov} and \eqref{eq:sm-shadow-covariance}, respectively. All additional moments $\langle PQ\rangle$ needed for off-diagonal entries are evaluated on the encoded state. Define the variance-dominated sample complexities
\begin{align}
 \widetilde n_{\rm cs}&=\frac{\log(1/\delta)}{\eta^2}
 \frac{1}{\boldsymbol{s}^{T}\Gamma^{+}\boldsymbol{s}},\nonumber\\
 \widetilde n_{\rm dir}&=\frac{\log(1/\delta)}{\eta^2}
 \frac{M}{\boldsymbol{s}^{T}V^{+}\boldsymbol{s}}.
 \label{eq:numerical-sample-complexities}
\end{align}
All numerical values of $\widetilde n_{\rm cs}$ and
$\widetilde n_{\rm dir}$ reported below use
$\eta=0.1$ and $\delta=0.05$. The pseudoinverse notation specifies the mathematical extension to singular cases; the reported numerical pipeline uses ordinary double-precision inversions and discards failed or nonfinite results. It applies no regularization and learns no covariance from synthetic measurement data.

The main plot uses the dimensionless coefficient
\begin{equation}
 R_K(\tau,N)={N\over3\mathcal I_K(\tau,N)}
 ={\widetilde n_{\rm cs}(\tau,N)\over\widetilde n_{\rm CSS}(N)},\qquad
 \widetilde n_{\rm CSS}(N)={3\log(1/\delta)\over\eta^2N}.
\end{equation}
Thus $R_K$ is also the ratio of local estimator variances at the same shot count. It removes the chosen error/confidence normalization and makes the separable threshold $R_K=1$ explicit. Values below one demonstrate the entanglement witness; the numerical calculations do not constitute a finite-shot experimental certification.

The weak dependence of the shadow results on $K$ should be distinguished from
the worst-case $3^K$ shadow-norm cost.  When the observable set is enlarged,
the old GLS solution, padded by zero coefficients for the newly added
observables, remains feasible.  The exact covariance-aware optimum can
therefore ignore directions that add no independent local information and
cannot become worse solely because more observables are made available.  The
small separation between the $K=2$ and $K=3$ curves in
Fig.~1 of the main text is consistent with this mechanism.

\subsection{Optimization over the twisting time}

The sample complexity is evaluated on a discrete grid of 22 twisting times for
every $(N,K)$.  For Fig.~1(a) of the main text, the row at $N=12$ is
shown over $0\leq\tau\leq0.2$ and normalized by the CSS value.  At each
$\tau$, the three cutoffs are evaluated on the same probe state.  The minimum
of each curve is refined by fitting a quadratic through the lowest grid point
and its two neighbors.  Its vertex is used only when the fitted curvature is
positive and the vertex lies within the local interval; otherwise the lowest
sampled point is retained.

Figure~1(b) of the main text uses the refined shadow minimum of $R_K$ for each $(N,K)$ within $0\le\tau\le0.2$. At fixed $N$, minimizing $R_K$ is equivalent to minimizing $\widetilde n_{\rm cs}$. The main comparison uses $4\le N\le12$. Shadow and direct sample complexities are optimized separately because their covariance matrices generally favor different probe states.
\section{Additional numerical results}
\label{app:additional-numerics}

This appendix collects numerical scans complementing Fig.~1 of the main text, using the sample complexities defined in Eq.~\eqref{eq:numerical-sample-complexities} and the same numerical parameters and calibration procedure.

\subsection{Resolved OAT optimization landscape}

\begin{figure*}[t]
  \centering
  \includegraphics[width=\textwidth]{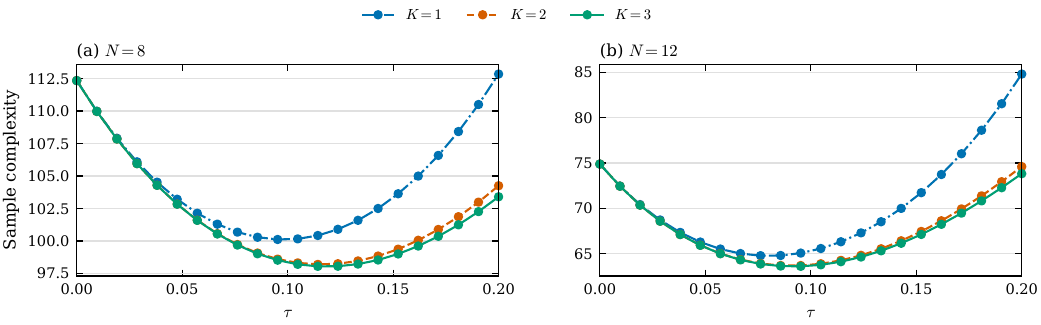}
  \caption{Shadow sample complexity $\widetilde n_{\rm cs}$ as a function of the OAT twisting time
  $\tau$ over the interval containing the optimized $\tau$, for
  (a) $N=8$ and (b) $N=12$.  The three curves correspond to moment cutoffs
  $K=1,2,3$.}
  \label{fig:oat-tau-zoomed}
\end{figure*}

Figure~\ref{fig:oat-tau-zoomed} resolves the shadow sample complexity over the
small-twisting-time interval $0\leq\tau\leq0.2$ for $N=8$ and $N=12$.  The
curves show that the minima used in the optimization are broad and lie
in the $\tau$ interval for all three moment cutoffs shown.  Increasing $K$ lowers
the sample complexity while producing only a modest displacement of the
minimum.  Thus, the improvement from higher-order moments in this regime does
not rely on selecting a qualitatively different OAT state.

\begin{figure}[t]
  \centering
  \includegraphics[width=0.72\columnwidth]{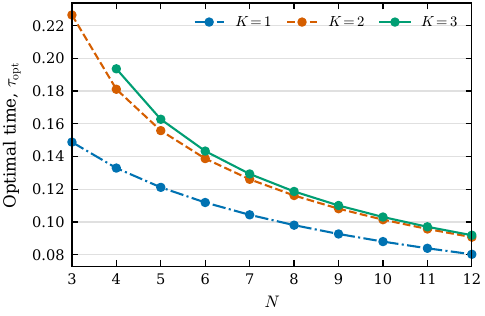}
  \caption{OAT twisting time $\tau_{\rm opt}$ that minimizes the shadow sample
  complexity as a function of $N$, for $K=1,2,3$.}
  \label{fig:oat-tau-opt}
\end{figure}

The corresponding optimal times are shown as a function of system size in
Fig.~\ref{fig:oat-tau-opt}.  Over the simulated range, $\tau_{\rm opt}$
decreases with $N$ for every cutoff.  The three sequences remain close,
consistent with the resolved scans in Fig.~\ref{fig:oat-tau-zoomed}. These finite-size data
describe the numerical optimization only; they do not by themselves
establish an asymptotic law for $\tau_{\rm opt}(N)$.

\subsection{Separate direct-measurement benchmark}

\begin{figure*}[t]
  \centering
  \includegraphics[width=\textwidth]{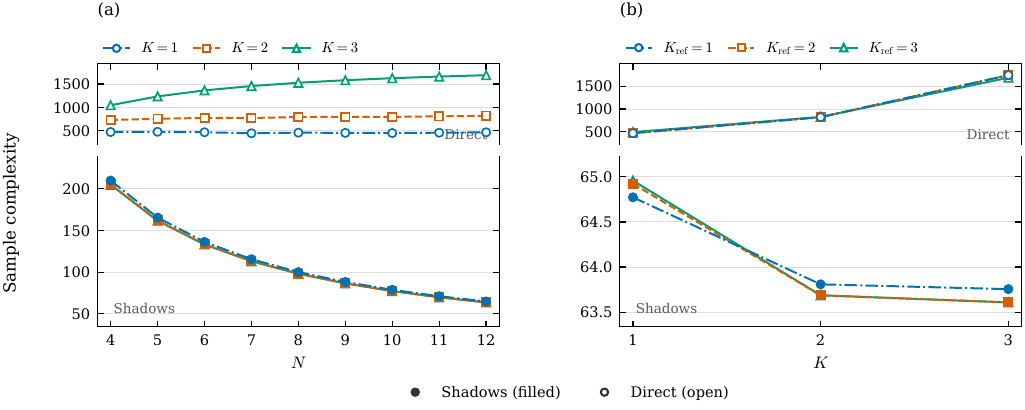}
  \caption{Sample complexity for OAT probes using shadows (filled markers) and separate, equally allocated direct measurements (open markers). (a) Dependence on $N$, with the twisting time optimized separately for every cutoff $K$ and protocol. (b) Dependence on $K$ at $N=12$: each color and shape identifies a reference cutoff $K_{\rm ref}$. For each protocol separately, its optimal twisting time at $K_{\rm ref}$ is held fixed across the three displayed cutoffs. Blue circles, orange squares, and green triangles denote cutoff values $1,2,3$, respectively. Lines connect data points as guides to the eye. Upper and lower axes display direct and shadow costs on separate vertical scales. Panel (a) uses $4\le N\le12$.}
  \label{fig:direct-benchmark}
\end{figure*}

For completeness, Fig.~\ref{fig:direct-benchmark} compares the shadow
protocol with the separate, equally allocated direct-measurement strategy
defined in the main text.  The direct benchmark assigns an independent shot
pool to every Pauli string and is therefore deliberately simple; it is not an
optimized direct protocol and should not be interpreted as a fundamental
limit on tailored or grouped measurements.  Its purpose is to isolate the
cost of foregoing data reuse.  Within this accounting, the direct complexity
grows as the observable set is enlarged, whereas the covariance-aware shadow
complexity remains nearly unchanged because GLS can suppress uninformative
directions.

\subsection{Twin-Fock probes}

Twin-Fock probes give a symmetry-distinct example of the effect of increasing
the maximum Pauli weight.  We use the $N$-qubit Dicke state
$\ket{D_N^{N/2}}$ and encode the parameter through
$\exp(-i\theta J_y)$ at the same working point $\theta_0=-0.5$.  Its
weight-two Pauli sector contains the components required to construct a
quadratic collective observable that can locally saturate the quantum Fisher
information~\cite{GessnerPRL2019,VolkoffFRONTIERS2024}.  Estimating those
components through randomized local Pauli measurements is not statistically
equivalent to directly implementing the corresponding collective
measurement, so higher-weight moments can still reduce the shadow-estimator
variance.

The equality of the $K=2$ and $K=3$ results in
Table~\ref{tab:twin-fock} follows from the symmetries of this protocol.  The
Twin-Fock state is an eigenstate of a collective $\pi$ rotation about $y$, a
symmetry preserved by the encoding, and the encoded state is real in the
computational basis.  Together these properties force every odd-weight Pauli
expectation and its derivative with respect to $\theta$ to vanish.  The shadow
covariance consequently separates into even- and odd-weight sectors, so
weight-three moments supply neither direct sensitivity nor useful covariance
with the even sector. Weight-four moments are not redundant: they lower the
complexity by approximately $22\%$ for $N=4$ and $7\%$ for $N=6$.

\begin{table}[t]
 \caption{Covariance-aware shadow-GLS sample complexity $\widetilde n_{\rm cs}$ for the
 $N$-qubit Twin-Fock probe $\ket{D_N^{N/2}}$ under the collective encoding
 $\exp(-i\theta J_y)$, evaluated at $\theta_0=-0.5$, as a function of the
 maximum Pauli weight $K$.  The target error and failure probability are
 $\eta=0.1$ and $\delta=0.05$, respectively.  The identical $K=2$ and $K=3$
 results reflect the decoupling of the even and odd-weight sectors.}
 \label{tab:twin-fock}
 \centering
 \begin{tabular}{cccc}
  \toprule
  $N$ & $K=2$ & $K=3$ & $K=4$ \\
  \midrule
  $4$ & $241.33$ & $241.33$ & $187.86$ \\
  $6$ & $177.88$ & $177.88$ & $165.24$ \\
  \bottomrule
 \end{tabular}
\end{table}

\end{document}